\documentclass[aps,pre,reprint,superscriptaddress,10pt]{revtex4-2}
\usepackage{amsmath, amssymb, graphicx, color, braket}
\usepackage{bbold}
\usepackage{verbatim}
\begin{document}

\title{Spectral and transport properties of a half-filled Anderson impurity coupled to phase-biased superconducting and metallic leads}

\author{Peter Zalom}
\email{zalomp@fzu.cz}
\affiliation{Institute of Physics, Czech Academy of Sciences, Na Slovance 2, CZ-18221 Praha 8, Czech Republic}

\author{Vladislav Pokorn\'y}
\email{pokornyv@fzu.cz}
\affiliation{Institute of Physics, Czech Academy of Sciences, Na Slovance 2, CZ-18221 Praha 8, Czech Republic}

\author{Tom\'a\v{s} Novotn\'y}
\email{tno@karlov.mff.cuni.cz}
\affiliation{Department of Condensed Matter Physics, Faculty of Mathematics and Physics, Charles University, Ke Karlovu 5, CZ-12116  Praha 2, Czech Republic}

\date{\today}

\begin{abstract}
We derive and apply a general scheme for mapping a setup consisting of a half-filled single level 
quantum dot coupled to one normal metallic and two superconducting phase-biased leads onto an ordinary 
half-filled single impurity Anderson model with single modified tunneling density of states. The theory 
allows for the otherwise unfeasible application of the standard numerical renormalization group and 
enables to obtain phase-dependent local spectral properties as well as phase-dependent induced 
pairing and Josephson current. The resulting transport properties match well with the numerically 
exact continuous-time hybridization-expansion quantum Monte Carlo. For weakly coupled normal electrode, 
the spectral properties can be interpreted in terms of normal-electrode-broadened Andreev bound states 
with phase-dependent position analogous to the superconducting Anderson model, which coexist in the 
$\pi$-like phase with a Kondo peak whose phase-dependent Kondo temperature is extracted. 

\end{abstract}

%\keywords{}

\maketitle


\section{Introduction \label{Intro} }

Gradual advance in experimental techniques over the past decades allowed to study electronic 
transport in increasingly sophisticated nanoscale systems with various, competing correlations. 
A prototype experiment typically includes a strongly interacting mesoscopic system attached to 
a reservoir with well defined properties. In theory, the mesoscopic system is frequently described 
in terms of one or multiple quantum dots (QDs) in the Coulomb blockade regime while the reservoir 
consists of normal metallic and/or  superconducting leads. Experimental realizations of such QDs include, 
for example, carbon nanotubes \cite{Tans-1997, Kasumov-1999, Jarillo-2006, Jorgensen-2006, Cleuziou-2006, Eichler-2009, Pillet-2010, Maurand-2012, Pillet-2013, Delagrange-2015} 
or semiconductor nanowires \cite{vanDam-2006, Lee-2012, Lee-2016, Xu-2017}.

The case of normal metallic electrodes attached to one QD can be modeled microscopically by the single impurity Anderson model (SIAM) which is one of the most understood models in the many-body physics \cite{Hewson-1993}. Here, free conduction electrons of the reservoir can completely or partially screen the magnetic doublet of QD depending on the parameters under the study. When the screening is effective, an emergent Kondo singlet becomes the ground state of the system \cite{Hewson-1993}. The Kondo singlet is a coherent many-body state %composed of a screening cloud and QD states. It 
with logarithmic energy scaling that can only be fully understood by applying renormalization group (RG) techniques \cite{Wilson-1975, Kopietz-2010, Kopietz-2013} or at least effective renormalization schemes \cite{Hewson-RG-2013,Janis-2007}. 

Once superconducting correlations are considered in the reservoir, the electron transport is altered by the Andreev scattering on the interface between the leads and QD \cite{Buitelaar-2002,Eichler-2007,Sand-Jespersen-2007,Buizert-2007,Grove-2007},
but it is still well described theoretically in terms of the Anderson impurity model with superconducting leads \cite{Luitz-2012}. For purely superconducting reservoirs, 
both the theoretical~\cite{Rodero-2011,Meden-2019} and experimental~\cite{Wernsdorfer-2010} understanding is fairly complete. In particular, a sufficiently large gap depopulates electrons around the Fermi energy to such an extent that the screening cloud around the impurity becomes disrupted. The ground state of the system changes then from a singlet (effective screening at small sized gaps) to a doublet, which is an example of an impurity quantum phase transition (QPT). This so-called $0$-$\pi$ transition is  accompanied  by  the  reversal  of  the  supercurrent, which is positive in the singlet and negative in the doublet phase \cite{vanDam-2006,Cleuziou-2006,Jorgensen-2007}. At the transition, one also observes the crossing of the Andreev bound states (ABSs) at the Fermi energy. %The complete phase diagram depends on the level energy $\varepsilon_d$, Coulomb repulsion $U$, hybridization strength of superconducting leads $\Gamma_S$ and on the superconducting phase difference across the junction, for details see Refs.\,\cite{Oguri-2012,Tanaka-2007}. %Depending on the relative strength of the on-dot Coulomb interaction the $0$-phase ground-state singlet can be predominantly BCS-like (for weak interaction) or Kondo-like (strong correlations) with a broad crossover between these two limiting cases. This physical picture has been firmly established over  the  years  by  various  analytic  and  numeric calculations.

Hybrid systems incorporating simultaneously normal as well as superconducting reservoirs, lead to even more intricate interplay of quantum correlation effects, where the understanding is limited both theoretically and experimentally \cite{Rodero-2011,Zitko-2015,Paaske-2015,Jellinggaard-2016,Domanski-2017,Satori-1992, Yoshioka-2000, Tanaka-2007, Oguri-2012, Oguri-2013}. In a simplest realization, one metallic and one superconducting lead have been studied experimentally in N-QD-S heterostructures \cite{Jellinggaard-2016}. From the theoretical perspective, such a problem is of only two-channel nature and thus well tractable by the numerical renormalization group (NRG) \cite{Zitko-2015} which offers unbiased insights and thus complements the purely numerical quantum Monte Carlo (QMC) simulations with reliable spectral properties \cite{Domanski-2017}. However, having just one superconducting lead does not allow superconducting phase difference across the QD. Consequently, such systems lack any supercurrent flow and no interplay of Kondo and Josephson effects takes place. Thus, the more interesting scenario includes one normal and two phase-biased superconducting leads. The resulting three-terminal structure is, however, beyond the reach of standard NRG since the corresponding discretized reservoir corresponds to three spin dependent and mutually interconnected hopping chains. The standard NRG scheme has thus been so far employed only to two channel problems \cite{Satori-1992, Yoshioka-2000}.

In the standard, computationally intractable approach to NRG \cite{Bulla-Rev-2008}, one would first discretize the bath of the three terminals. The resulting semi-infinite hopping chain would be then transformed via the Bogolyubov-Valatin transformation at each chain site. Subsequently, in the  second step, particle-hole transformations on odd and even sites separately need to be carried out \cite{Satorni-1992, Yoshioka-2000}. The resulting Wilson chain would, however, consist of three mutually interconnected spin-polarized chains which is beyond the present computational power \cite{Domanski-2017}. To circumvent the problem, one may apply a general procedure of Ref.~\cite{Liu-2016} or introduce suitable unitary transformations to diagonalize the problem \cite{Tanaka-2007, Oguri-2012, Oguri-2013}. The second approach has already been successfully applied to the hybrid normal-superconductor reservoir problem in the limit of infinite superconducting gap. Although the authors of Refs.~\cite{Tanaka-2007, Oguri-2012, Oguri-2013} make explicit reference to tnhe Wilson chains corresponding to discretized versions of the model under the study, as shown in the present paper, it is possible to limit the transformations in the case of half-filling \footnote{In the present model, half-filling implies that the system is tuned to the particle-hole symmetric point, see Ref.~\cite{Tanaka-2007}.} to just the local electrons of QD and map the infinite-gap model onto an ordinary asymmetric SIAM directly. 

Generalizing the approach of Refs.~\cite{Tanaka-2007, Oguri-2012, Oguri-2013}, we are able to treat the present general three-terminal problem at the half-filling with finite superconducting gap and map it onto a single impurity Anderson model of fermions with tunneling density of states (TDOS) in the reservoir that corresponds to the  standard one-channel-lead case tractable by NRG in the scheme of Ref.~\cite{Bulla-1994}. Since it is believed that such an approach is not feasible \cite{Hecht-2008}, we present the details of the transformation in Sec.~\ref{sec_2} where also a detailed microscopic formulation of the problem is stated. In Sec.~\ref{sec_3}, we proceed to the $\Delta \rightarrow \infty$ case treated previously in Refs.~\cite{Tanaka-2007, Oguri-2012, Oguri-2013} and show that their approach is completely equivalent with ours when the half-filled case is considered. However, as opposed to Refs.~\cite{Tanaka-2007, Oguri-2012, Oguri-2013}, no reference to NRG discretization is required. Finally, in Sec.~\ref{sec_4}, the general three-terminal problem with finite superconducting gap is solved at the half-filling. To this end, the mapping of the finite-gap problem onto a single-channel SIAM with altered TDOS is performed. Subsequently, standard NRG approach of Ref.~\cite{Bulla-1994} is employed utilizing the NRG Ljubljana code \cite{Ljubljana-code}. Using the backwards transformations, all spectral and transport properties of the original three-terminal setup are then determined. The most important conclusions are summarized in Sec.~\ref{sec_5}. Technical calculations regarding the transformation of the interaction term in Sec.~\ref{sec_2} and the effect of the finite band width are discussed in the Appendices~\ref{app_hubbard} and \ref{app_finite}, respectively. The comparison with QMC is shown in the Appendix~\ref{app_qmc}.

\section{Mapping onto SIAM-like models \label{sec_2}}

\subsection{Microscopic formulation \label{subsec_micro}}

\begin{figure*}
	\begin{center}
		\includegraphics[width=1.60\columnwidth]{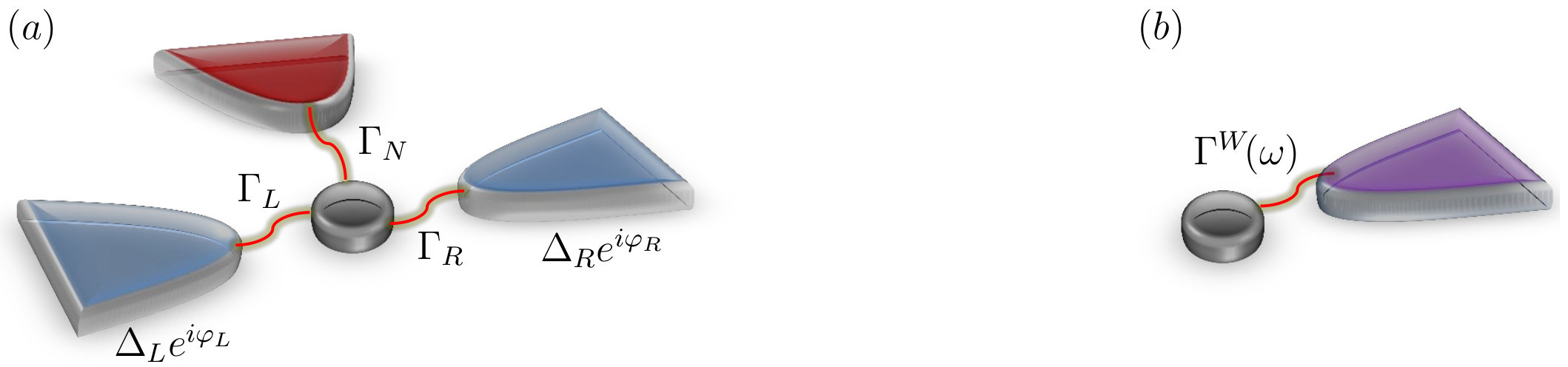}
		\vspace{0.2cm}
		\caption{$(a)$ Scheme of $Y$-shaped three-terminal junction where a QD (black disc) couples via the hybridization strength $\Gamma_N$ to one normal electrode (red  pointed teardrop) and left (L) and the right (R) superconducting electrode (blue  pointed teardrop) which are phase-biased by $\varphi = \varphi_L -\varphi_R$. The BCS leads hybridize with quantum dot with strengths $\Gamma_L$ and $\Gamma_R$, respectively, and are considered here as having the same BCS gap parameter $\Delta_L = \Delta_R \equiv \Delta$.  
		$(b)$ Equivalent scheme consisting of one-terminal reservoir containing  Bogolyubov-like quasiparticles (purple  pointed teardrop) which are hybridized with the QD via a structured hybridization function $\Gamma^W(\omega)$. In Sec.~\ref{subsec_transform} we show that by employing $\Gamma^W(\omega)$ according to Eq.~(\ref{TDOS_WB}) the schemes $(a)$ and $(b)$ may be in terms of spectral properties mapped onto each other at the half-filling. To obtain transport properties corresponding to the $Y$ shape geometry of panel $(a)$ transformations according to Sec.~\ref{subsec_transp} are required. \label{fig_1}}
	\end{center}
\end{figure*}

The hybrid three-terminal setup consists of a mesoscopic system modeled as a usual Anderson magnetic impurity connected to one normal metallic and two superconducting electrodes. The superconducting electrodes follow the Bardeen-Cooper-Schrieffer (BCS) theory with one lead referred to as the left ($L$) and the other one as the right ($R$), see Fig.~\ref{fig_1}. The total Hamiltonian of the system is then the sum of the dot Hamiltonian $H_{d}$, the Hamiltonian of the normal lead $H_N$, two BCS Hamiltonians for superconducting leads $H_L$ and $H_R$, and three tunneling Hamiltonians $H_{T,\alpha}$ with $\alpha \in \{N, L, R\}$ which connect each lead separately to the dot. The constituent Hamiltonians read as 
\begin{eqnarray}
H_d
&=&
\sum_{\sigma} 
\varepsilon_{d}
d^{\dagger}_{\sigma}
d^{\vphantom{\dagger}}_{ \sigma}
+
U
d^{\dagger}_{\uparrow}
d^{\vphantom{\dagger}}_{ \uparrow}
d^{\dagger}_{\downarrow}
d^{\vphantom{\dagger}}_{ \downarrow},
\label{dotH}
\\
H_{\alpha}  
&=&
\sum_{\mathbf{k}\sigma} 
\, \varepsilon_{\mathbf{k}\alpha}
c^{\dagger}_{\alpha \mathbf{k} \sigma}
c^{\vphantom{\dagger}}_{\alpha \mathbf{k} \sigma}
\nonumber
\\
&-&
\Delta_{\alpha}
\sum_{\mathbf{k}}
\left(e^{i\varphi_\alpha} 
c^{\dagger}_{\alpha \mathbf{k} \uparrow} 
c^{\dagger}_{\alpha -\mathbf{k} \downarrow}
+
\textit{H.c.}\right),
\label{kineticH}
\\
H_{T,\alpha} 
&=&
\sum_{\mathbf{k} \sigma} \,
\left(V^*_{\alpha\mathbf{k}} 
c^{\dagger}_{\alpha\mathbf{k}\sigma}
d^{\vphantom{\dagger}}_{\sigma} 
+ 
V_{\alpha\mathbf{k}} 
d^{\dagger}_{\sigma}
c^{\vphantom{\dagger}}_{\alpha\mathbf{k}\sigma}\right),
\label{tunnelH}
\end{eqnarray}
where $c^{\dagger}_{\alpha \mathbf{k} \sigma}$ creates an electron of spin $\sigma \in \{\uparrow \downarrow \}$ and quasi-momentum $\mathbf{k}$ in the lead $\alpha$ while $c^{\vphantom{\dagger}}_{\alpha\mathbf{k}\sigma}$ annihilates it. In analogy, $d^{\dagger}_{\sigma}$ creates a dot electron of spin $\sigma$ while $d^{\vphantom{\dagger}}_{\sigma}$ annihilates it. The QD is characterized by the Coulomb repulsion $U$ and the level energy $\varepsilon_d$ which in the most general case is arbitrary but we will later concentrate only at $\varepsilon_d=-U/2$. The QD hybridizes with the leads via $V_{\alpha\mathbf{k}}$ and the gap parameter vanishes in the normal lead, thus $\Delta_N=0$. 

In all our calculations we use dimensionless units with $\hbar=1$ and $e=1$. Moreover, we focus on a generic case with a constant TDOS with a finite half-bandwidth $B$ 
\begin{equation}\label{TDOS}
\Gamma_\alpha(\omega)=\pi\sum_{\mathbf{k}}|V_{\alpha\mathbf{k}}|^2\delta(\omega-\varepsilon_{\mathbf{k}\alpha})
=\Gamma_\alpha\Theta(B^2-\omega^2)\ ,
\end{equation} 
which also defines the tunneling rates $\Gamma_\alpha$.
We concentrate on the case with symmetric coupling to the superconducting leads $\Gamma_L=\Gamma_R\equiv\Gamma_S/2$
as any asymmetric case $\Gamma_L\neq\Gamma_R$ can be obtained from the symmetric one using the procedure
described in Ref.~\cite{Kadlecova-2017}.

We also restrict to situations with the same gap parameters in both 
superconducting leads $\Delta_L=\Delta_R\equiv\Delta$ as this is a typical situation in an experiment. Concerning the BCS phase parameters $\varphi_L$ and $\varphi_R$, as in any Josephson junction physical observables
can only depend on the phase difference $\varphi=\varphi_L-\varphi_R$ and not on their individual values, i.e.~they must be invariant with respect to a global phase shift $\varphi_{L,R}\rightarrow\varphi_{L,R}+\varphi_{\textrm{s}}$
which is a manifestation of the gauge invariance \cite{Meden-2019}. Therefore, we are free to choose a convenient symmetric phase-drop setup with $\varphi_L=-\varphi_R=\varphi/2$ in what follows.

%XXX
%To allow for an unbiased treatment of the Kondo and superconducting correlations on the same footing, RG techniques should be applied. However, in the basis of the $d$ and $c$ electrons, superconducting leads mix together the spin up and spin down electrons of the bath which requires to solve a general problem of three mutually interacting spin-polarized-bath channels. Applying the original numerical RG scheme \cite{Wilson-1975} is therefore extremely demanding from the point of view of the computational time \cite{Domanski-2017}. In general, multichannel problems pose a genuine challenge when applying NRG and have been tackled only quite recently within the interleaved formulation of the generalized Wilson chain \cite{Mitchell-2014}.

The Hamiltonians of all three leads are quadratic due to the standard non-interacting assumption. Consequently, the lead electrons can be integrated out to obtain a genuine one channel impurity problem. To this end, it is advantageous to reformulate Eqs.~(\ref{dotH})-(\ref{tunnelH}) using the Nambu formalism.

\subsection{Hamiltonian in the Nambu basis \label{subsec_nambu} }

Nambu formalism represents a convenient way of re-arranging Hamiltonians involving BCS superconductivity in a way where all the lead electrons are treated on an equal footing. To this end, let us combine the spin up and spin down component of the corresponding fields describing the $c$ electrons into spinors
\begin{equation}
C^{\dagger}_{\alpha \, \mathbf{k}} 
=
\left(
c^{\dagger}_{\alpha \, \mathbf{k} \uparrow},
c^{\vphantom{\dagger}}_{\alpha \, -\mathbf{k} \downarrow}
\right)
\end{equation}
with $\alpha \in \{N,L,R\}$, while the spinor $D$ of the dot electrons $d$ is constructed in complete analogy as
\begin{equation}
D^{\dagger} 
=
\left(
d^{\dagger}_{\uparrow},
d^{\vphantom{\dagger}}_{\downarrow}
\right).
\end{equation}
Under the standard BCS assumption $\varepsilon_{\mathbf{k}\alpha}=\varepsilon_{-\mathbf{k}\alpha}$ and with a convenient choice of real tunnel couplings $V_{\alpha\mathbf{k}}=V^*_{\alpha\mathbf{k}}=V_{\alpha-\mathbf{k}}$,  the Hamiltonians (\ref{kineticH}) and (\ref{tunnelH}) apart from possible unimportant constant energy shifts then become 
\begin{eqnarray}
H_{\alpha} 
&=&
\sum_{\mathbf{k}} 
C_{\alpha\, \mathbf{k}}^{\dagger} 
\mathbb{E}^{\vphantom{\dagger}}_{\alpha \, \mathbf{k}}
C_{\alpha\, \mathbf{k}}^{\vphantom{\dagger}},
\label{NambuHalpha}
\\
H_{T,\alpha}
&=&
\sum_{\mathbf{k}}
\left(
D^{\dagger}
\mathbb{V}^{\vphantom{\dagger}}_{\alpha \, \mathbf{k}} 
C_{\alpha\mathbf{k}}^{\vphantom{\dagger}}
+
C_{\alpha\mathbf{k}}^{\dagger} 
\mathbb{V}^{\vphantom{\dagger}}_{\alpha \, \mathbf{k}} 
D^{\vphantom{\dagger}}
\right)
\label{NambuHtunel}
\end{eqnarray}
with
\begin{eqnarray}
\mathbb{E}_{\alpha\mathbf{k}}  
& = &
-\Delta_\alpha C_{\alpha} \sigma_x
+\Delta_\alpha S_{\alpha} \sigma_y
+ \varepsilon_{\mathbf{k}\alpha} \sigma_z,
\label{Ealpha}
\\
\mathbb{V}_{\alpha\mathbf{k}}
& = &
V_{\alpha\mathbf{k}}  \, \sigma_z,
\label{Valpha}
\end{eqnarray}
where  $\sigma_i$, $i\in \{x,y,x\}$, are the Pauli matrices 
while $C_{\alpha} \equiv \cos{\varphi_\alpha}$, $S_{\alpha} \equiv \sin{\varphi_\alpha}$. The blackboard bold typeface is from now on used to distinguish matrices from scalars.

Before applying the Nambu formalism to the Hamiltonian (\ref{dotH}), let us first separate it into a quadratic part
\begin{eqnarray}
H_{d,0}
&=&
\sum_{\sigma}\varepsilon_d
d^{\dagger}_{\sigma}
d^{\vphantom{\dagger}}_{ \sigma}
+\frac{U}{2}
D^{\dagger} 
\left(
\sigma_x+\sigma_z
\right)
D^{\vphantom{\dagger}}
\nonumber
\\
&=&
D^{\dagger} 
\mathbb{E}_d 
D^{\vphantom{\dagger}}
\label{NambuHdot}
\end{eqnarray}
with
\begin{equation}
\mathbb{E}_d  
=
\frac{U}{2} \sigma_x
+
\left(\frac{U}{2} + \varepsilon_{d} \right)
\sigma_z
\label{Ed}
\end{equation}
and a mixed quadratic and quartic interaction part
\begin{equation}
H_U
=
U
d^{\dagger}_{\uparrow}
d^{\vphantom{\dagger}}_{ \uparrow}
d^{\dagger}_{\downarrow}
d^{\vphantom{\dagger}}_{ \downarrow}
-\frac{U}{2}
D^{\dagger} 
\left(
\sigma_x+\sigma_z
\right)
D^{\vphantom{\dagger}}.
\label{InteractionD}
\end{equation}
The advantage of this non-standard partitioning will be discussed in Sec.~\ref{subsec_transform}. %For now, let us merely state that although under such partitioning the interaction part does acquire additional quadratic terms, the non-interacting part $H_{d,0}$ remains just quadratic. 

Taking together,  in the Nambu formalism the non-interacting quadratic part of the present problem spanned by the $D$ and $C$ spinors reads as
\begin{equation}
H_0
=
H_{d ,0}
+
\sum_{\alpha}
\left(
H_{\alpha}
+
H_{T,\alpha}
\right)
\label{H_0}
\end{equation}
while the modified interaction part $H_U$ is given by Eq.~(\ref{InteractionD}).

\subsection{Bogolyubov-Valatin transformations in the space of local electrons \label{subsec_transform}}

Let us now study the effect of unitary transformations $\mathbb{T}$ and $\tilde{\mathbb{T}}_{\alpha}$ onto the spinors $D$ and $C_{\alpha}$ respectively. We introduce the spinors $W$ and $\tilde{C}$ via
\begin{eqnarray}
\mathbb{T} D
&\equiv&
W,
\qquad  \, \, \, \, \, \, \, \,
D^{\dagger} \mathbb{T}^{\dagger}
\equiv
W^{\dagger},
\\
\tilde{\mathbb{T}}_{\alpha} C_{\alpha\mathbf{k}}
&\equiv&
\tilde{C}_{\alpha\mathbf{k}},
\qquad 
C^{\dagger}_{\alpha\mathbf{k}}
\tilde{\mathbb{T}}^{\dagger}_{\alpha}
\equiv
\tilde{C}^{\dagger}_{\alpha \mathbf{k}}.
\label{TransformedW}
\end{eqnarray}

At this point, we consider no other constraints on the transformations $\mathbb{T}$, $\tilde{\mathbb{T}}_{\alpha}$  except of unitarity so that the many-body energy spectra of the problem remain the same in both spinor bases $D$ and $W$. On the other hand, such transformations may crucially affect the form of the one-particle operators in the corresponding non-interacting Hamiltonians thus allowing for computationally more suitable non-interacting Green functions and/or self-energy contributions from the integrable degrees of freedom for the problem under the study. 

The effect of the transformations $\mathbb{T}$ and $\tilde{\mathbb{T}}_{\alpha}$ on the quadratic part $H_{d,0}$ reads as
\begin{equation}
H_{d,0}
= 
W^{\dagger} 
\left(\mathbb{T} \mathbb{E}_d \mathbb{T}^{\dagger}\right)
W^{\vphantom{\dagger}},
\end{equation}
while the tunneling Hamiltonians change as
\begin{eqnarray}
H_{T,\alpha}
&=&
\sum_{\mathbf{k}}
W^{\dagger}
\left(
\mathbb{T}
\mathbb{V}^{\vphantom{\dagger}}_{\alpha\mathbf{k}}
\tilde{\mathbb{T}}_{\alpha}^{\dagger}
\right)
\tilde{C}_{\alpha\mathbf{k}}^{\vphantom{\dagger}}
\nonumber
\\
&+&
\sum_{\mathbf{k}}
\tilde{C}^{\dagger}_{\alpha\mathbf{k}}
\left(
\tilde{\mathbb{T}}_{\alpha}
\mathbb{V}^{\vphantom{\dagger}}_{\alpha\mathbf{k}}
\mathbb{T}^{\dagger}
\right)
W.
\end{eqnarray}

The kinetic Hamiltonians are affected only by the transformations $\tilde{\mathbb{T}}_{\alpha}$ as they involve no operators of the local electrons:
\begin{equation}
H_{\alpha}
= 
\sum_{\mathbf{k}} 
\tilde{C}_{\alpha\, \mathbf{k}}^{\dagger} 
\left(
\tilde{\mathbb{T}}_{\alpha}
\mathbb{E}^{\vphantom{\dagger}}_{\alpha\mathbf{k}}
\tilde{\mathbb{T}}_{\alpha}^{\dagger}
\right)
\tilde{C}_{\alpha\, \mathbf{k}}^{\vphantom{\dagger}}.
\end{equation}

Since the non-interacting ($U=0$) Hamiltonian is quadratic  we can easily obtain the retarded Green function $\mathbb{G}_0^W(\omega^+)$ \footnote{The argument $\omega^+$ of the function emphasizes to which part of the complex $z$-plain it belongs (above vs. below the real axis of $z$). This notation is employed throughout this paper when required.} which corresponds to the $W$ spinors. Employing the equation of motion technique for the Green functions in an exact analogy to Ref.~\cite{Novotny-Rossini-2005}, we introduce an infinite-dimensional vector
\begin{equation}
\Psi^{\dagger}
=
(W^{\dagger},~\tilde{C}^{\dagger}_{N\mathbf{k}},~\tilde{C}^{\dagger}_{L\mathbf{k}},~\tilde{C}^{\dagger}_{R\mathbf{k}}),
\end{equation}
where $\Psi$ is its Hermitian conjugate and the spinors $\tilde{C}^{\dagger}_{\alpha\mathbf{k}}$ are understood to be repeated in $\Psi^{\dagger}$ for all possible quasi-momenta of lead electrons. This allows us to rearrange the non-interacting Hamiltonian as
\begin{equation}
H_0
=
\Psi^{\dagger} 
\mathbb{E}^W
\Psi^{\vphantom{\dagger}},
\end{equation}
with
\begin{eqnarray}
\mathbb{E}^W
=
\left(
\begin{matrix}
\mathbb{T} \mathbb{E}_d \mathbb{T}^{\dagger}
&
\mathbb{T} \mathbb{V}^{\vphantom{\dagger}}_{N \, \mathbf{k}} \tilde{\mathbb{T}}_N^{\dagger}
&
\mathbb{T} \mathbb{V}^{\vphantom{\dagger}}_{L \, \mathbf{k}} \tilde{\mathbb{T}}_L^{\dagger}
&
\mathbb{T} \mathbb{V}^{\vphantom{\dagger}}_{R \, \mathbf{k}} \tilde{\mathbb{T}}_R^{\dagger}
\\
\tilde{\mathbb{T}}_N \mathbb{V}^{\vphantom{\dagger}}_{L \, \mathbf{k}} \mathbb{T}^{\dagger}
&
\tilde{\mathbb{T}}_N \mathbb{E}^{\vphantom{\dagger}}_{N \, \mathbf{k}} \tilde{\mathbb{T}}_N^{\dagger}
&
0
&
0
\\
\tilde{\mathbb{T}}_L \mathbb{V}^{\vphantom{\dagger}}_{L \, \mathbf{k}} \mathbb{T}^{\dagger}
&
0
&
\tilde{\mathbb{T}}_L \mathbb{E}^{\vphantom{\dagger}}_{L \, \mathbf{k}} \tilde{\mathbb{T}}_L^{\dagger}
&
0
\\
\tilde{\mathbb{T}}_R \mathbb{V}^{\vphantom{\dagger}}_{R \, \mathbf{k}} \mathbb{T}^{\dagger}
&
0
&
0
&
\tilde{\mathbb{T}}_R \mathbb{E}^{\vphantom{\dagger}}_{R \, \mathbf{k}} \tilde{\mathbb{T}}_R^{\dagger}
\end{matrix}
\right),
\nonumber
\\
\label{E^W}
\end{eqnarray}
where the upper index $W$ was introduced to clearly distinguish the underlying spinor basis $W$ for the formulation of the infinite-dimensional matrix $\mathbb{E}^W$. 

In general, the non-interacting problem can be solved by finding the retarded Green function $\mathbb{G}_0(\omega^+)$  where $\omega^+ \equiv \omega +i\eta$ while $\omega$ is a real frequency and $\eta$ is an infinitesimally small positive number. To this end, standard equation of motion technique formulated in the matrix form requires one to solve the resolvent equation $\mathbb{G}_0(\omega^+)=\left(\omega^+ \mathbb{1}-\mathbb{H}_0\right)^{-1}$ with $\mathbb{1}$ being the unit matrix. However, for the present problem we only need to obtain the solution for the local Green function of the dot electrons which corresponds the left upper $2 \times 2$ block of expression (\ref{E^W}). Employing the partitioning scheme of Ref.~\cite{Novotny-Rossini-2005}, we obtain the local retarded Green function $\mathbb{G}^W_0(\omega^+)$ in the spinor basis $W$ directly as 
\begin{eqnarray}
\mathbb{G}^W_0(\omega^+)
&=&
\left(
\omega^+ \mathbb{1}
- 
\mathbb{T} \mathbb{E}_d \mathbb{T}^{\dagger} 
- 
\mathbb{T} \mathbb{\Sigma}^D \mathbb{T}^{\dagger}
\right)^{-1}
\label{G_0^W}
\end{eqnarray}
with
\begin{eqnarray}
\mathbb{\Sigma}^D(\omega^+)
&=&
\sum_{\alpha \mathbf{k}} 
\mathbb{V}_{\alpha \mathbf{k}} 
\tilde{\mathbb{T}}_{\alpha}^{\dagger} 
\left(
\omega^+ \, \mathbb{1}
- 
\tilde{\mathbb{T}}_{\alpha} 
\mathbb{E}_{\alpha \mathbf{k}}
\tilde{\mathbb{T}}_{\alpha}^{\dagger}
\right)^{-1} 
\tilde{\mathbb{T}}_{\alpha} 
\mathbb{V}_{\alpha \mathbf{k}}
\nonumber
\\
&=&
\sum_{\alpha \mathbf{k}} 
\mathbb{V}_{\alpha \mathbf{k}} 
\left(
\omega^+ \, \mathbb{1}
- 
\mathbb{E}_{\alpha \mathbf{k}}
\right)^{-1} 
\mathbb{V}_{\alpha \mathbf{k}},
\label{SigmaD}
\end{eqnarray}
which thus represents the self-energy contribution from the leads expressed with respect to the spinor basis $D$ (as denoted by the upper index $D$). Moreover, one may also obtain the self-energy contribution $\mathbb{\Sigma}^W$ as
\begin{equation}
\mathbb{\Sigma}^W(\omega^+)
=
\mathbb{T}
\mathbb{\Sigma}^D(\omega^+)
\mathbb{T}^{\dagger},
\label{SigmaW}
\end{equation}
which not only defines $\mathbb{\Sigma}^W(\omega^+)$, but also gives us the transformation rule to easily interchange the spinor bases $D$ and $W$ when required. By exploiting the unitarity of $\mathbb{T}$, we may also extract an analogous transformation rule for the non-interacting retarded Green functions
\begin{equation}
\mathbb{G}^W_0(\omega^+)
=
\mathbb{T}
\left(
\omega^+ \mathbb{1}
- \mathbb{E}_d  
- \mathbb{\Sigma}^D 
\right)^{-1}
\mathbb{T}^{\dagger}
=
\mathbb{T}
\mathbb{G}_0^D(\omega^+)
\mathbb{T}^{\dagger}
\label{G_0^D}
\end{equation}
which yields the transformation rule as well as the definition of the non-interacting ($U=0$) retarded Green function with respect to the spinor basis $D$. Clearly, in both bases the effect of the leads is fully integrated out and only enters the $\mathbb{G}_0^W(\omega^+)$ and $\mathbb{G}_0^D(\omega^+)$
via the corresponding self-energy contributions $\mathbb{\Sigma}^W(\omega^+)$ and $\mathbb{\Sigma}^D(\omega^+)$, respectively. Green function as well as the self-energy contributions in different bases relate to each other via the local dot transformation $\mathbb{T}$ since the transformations $\tilde{\mathbb{T}}_{\alpha}$ are canceled out in Eqs.~(\ref{SigmaW}) and (\ref{G_0^D}). 

It is now our aim to construct a suitable transformation $\mathbb{T}$, so that the self-energy contribution $\mathbb{\Sigma}^W$ is diagonal. To this end, we first perform all summations in Eq.~(\ref{SigmaD}), which is a quite straightforward with details given in the Appendix~\ref{app_finite}. The resulting expression for $\mathbb{\Sigma}^D(\omega^+)$ has the following matrix structure:
\begin{equation}\label{eq:self-energy}
\mathbb{\Sigma}^D(\omega^+)
=
\Sigma^D_n(\omega^+) \mathbb{1} 
+ 
\Sigma_a^D (\omega^+) \mathbb{\sigma}_x,
\end{equation}
where $\Sigma^D_n(\omega)$ and $\Sigma^D_a(\omega)$ are functions of frequency with  the form (for here and now unimportant) which is given in the Appendix~\ref{app_finite}. We insert now $\mathbb{\Sigma}^D(\omega^+)$ back into Eq.~(\ref{G_0^D}) to obtain the matrix structure of $\left(\mathbb{G}_0^D\right)^{-1}$. We first concentrate exclusively on the half-filled case where $\left(\mathbb{G}_0^D\right)^{-1}$ has a much simpler structure. Afterwards, the more general $\varepsilon_d \neq -U/2$ case is inspected.  

At the half-filling the diagonal part of $\left(\mathbb{G}_0^D\right)^{-1}$ is only proportional to the unit matrix $\mathbb{1}$ which remains unaltered under unitary transformation. On the other hand, the off-diagonal parts of $\mathbb{\Sigma}^D$ and $\mathbb{E}_d$ are both proportional to $\sigma_x$ and may thus be simultaneously diagonalized by enforcing the condition $\mathbb{T} \sigma_x \mathbb{T}^{\dagger} = \pm \sigma_z$ onto the transformation $\mathbb{T}$. The resulting non-interacting Green function $\mathbb{G}_0^W$ is then also diagonal as intended. Each sign option of the condition $\mathbb{T} \sigma_x \mathbb{T}^{\dagger} = \pm \sigma_z$, is solved by two linearly independent transformations which we denote $\mathbb{T}_1^{\pm}$ and $\mathbb{T}_2^{\pm}$ with the given superscript indicating the sign of $\sigma_z$ in the condition. Explicitly, we obtain
\begin{eqnarray}
\mathbb{T}_1^{\pm}
&=
\frac{1}{\sqrt{2}}
\left(
\sigma_x \pm \sigma_z
\right),
\nonumber
\\
\mathbb{T}_2^{\pm}
&=
\frac{1}{\sqrt{2}}
\left(
\mathbb{1} \pm i \sigma_y
\right).
\label{Transformations}
\end{eqnarray}
All transformations fulfill $\mathbb{T}^{\dagger} = \mathbb{T}^{-1} = \mathbb{T}$ and are of Bogolyubov-Valatin type but without momentum or  frequency dependence. 

NOtably, trying to generalize the previous approach to $\varepsilon_d \neq -U/2$ changes the matrix structure of $\left(\mathbb{G}_0^D\right)^{-1}$ considerably by adding an extra diagonal term proportional to $(\varepsilon_d +U/2)\sigma_z$. Diagonalizing $\left(\mathbb{G}_0^D\right)^{-1}$ by $\mathbb{T}$ would now require to simultaneously fulfill not only $\mathbb{T} \sigma_x \mathbb{T}^{\dagger} = \pm \sigma_z$ but also $\mathbb{T} \sigma_z \mathbb{T}^{\dagger} = \pm \sigma_z$ or $\mathbb{T} \sigma_z \mathbb{T}^{\dagger} = \pm \mathbb{1}$. Neither of the two combined conditions is however solvable. Consequently, outside of the half-filled case there exists no unitary transformation to diagonalize $\left(\mathbb{G}_0^D\right)^{-1}$. For example, applying $\mathbb{T}$ in the form of Eq.~(\ref{Transformations}) away from the half-filling rotates the superconducting terms onto the diagonal elements of $\left(\mathbb{G}_0^W\right)^{-1}$ which is however traded off for rotating the originally diagonal terms proportional to $(\varepsilon_d +U/2)\sigma_z$ into off-diagonal terms proportional to $(\varepsilon_d +U/2)\sigma_x$. From now on, we therefore concentrate \textit{exclusively on the half-filled case}.

Although, for the non-interacting ($U=0$) half-filled case the corresponding Green function can be diagonalized by transformations (\ref{Transformations}), in the end, in the full interacting case the action of transformations (\ref{Transformations}) on the interaction part $H_U$ needs to be considered. As shown in the Appendix \ref{app_hubbard}, using the interaction term mixed of quadratic and quartic terms according to Eq.~(\ref{InteractionD}) allows to obtain the Hubbard interaction term when the transformation $\mathbb{T}_1^-$ is used. Explicitly:
\begin{equation}
H_U
=
U
w^{\dagger}_{\uparrow}
w^{\vphantom{\dagger}}_{ \uparrow}
w^{\dagger}_{\downarrow}
w^{\vphantom{\dagger}}_{ \downarrow},
\label{Hub_w}
\end{equation}
The remaining options for $\mathbb{T}$ produce all an extra quadratic terms in $H_U$ which is proportional to $\sigma_x$, thus spoiling the diagonal form of $\mathbb{G}_0^W(\omega^+)$. We thus select the transformation $\mathbb{T}_1^-$ in what follows and denote it as $\mathbb{T} \equiv \mathbb{T}^-_1$. Nevertheless, we stress that the remaining choices would also be possible when different partitionings of the full Hamiltonian are considered. Moreover, the transformation $\mathbb{T}_2^+$ fully corresponds to the transformation used in the standard NRG treatments of magnetic impurities coupled to superconducting reservoirs (see Refs.~\cite{Satori-1992,Yoshioka-2000} for more detail) while  $\mathbb{T}_1^+$ relates to the transformations applied in the Refs.~\cite{Tanaka-2007,Oguri-2012, Oguri-2013} to solve the $\Delta \rightarrow \infty$ limit of the present model. However, in all of the aforementioned references, even at the half-filling the lead transformations corresponding to $\tilde{\mathbb{T}}_{\alpha}$ are explicitly performed within the applied NRG algorithms.  

Since the resulting $\mathbb{G}_0^W(\omega^+)$ is proportional to $\omega \sigma_z g_1(\omega)+g_2(\omega)\mathbb{1}$ with $g_1(\omega)$ and $g_2(\omega)$ being even functions of $\omega$ we may actually drop the matrix Nambu formalism and employ directly the $w^{\dagger}_{\uparrow}$ and $w^{\dagger}_{\downarrow}$ fields which constitute the $W$ spinor. We stress out in this regard, that such a transition requires us to change the hole propagator $\mathbb{G}_0^W(\omega^+)$ to an electron propagator which involves simultaneous change of the frequency sign as well as one extra minus sign for normal ordering of the creation and annihilation operators to obtain the corresponding spin-down electron propagator of the constituent field $w^{\dagger}_{\downarrow}$. In detail, $\omega g_1(\omega) \sigma_z \rightarrow \omega g_1(\omega)\mathbb{1}$ and $g_2(\omega)\mathbb{1} \rightarrow g_2(\omega) \mathbb{1}$. Thus, one obtains
\begin{equation}
G_{0\uparrow}^W (\omega^+)
= 
G_{0 \downarrow}^W (\omega^+)
= 
\frac{1}{\omega + U/2 - \Sigma^W(\omega^+) },
\label{G_0w}
\end{equation}
and the independence of Green functions of the spin index of the field $w$ becomes explicit. The calculation of $\Sigma^W(\omega^+)=\Sigma^D_n(\omega^+)-\Sigma^D_a(\omega^+)$ can be performed for arbitrary bandwidth as shown in the Appendix~\ref{app_finite}. In the limit of infinitely wide band $B\to\infty$, it takes the following form 
\begin{eqnarray}
\Sigma^W(\omega^+)
&=&
-i\Gamma_N
-\Gamma_{S}
\left[
\frac{i~ \mathrm{sgn}(\omega) }{\sqrt{\omega^2-\Delta^2}}
\Theta(\omega^2-\Delta^2)%_{\textit{out}}
\right.
\nonumber
\\
&+& 
\left.
\frac{ \Theta(\Delta^2-\omega^2) }{\sqrt{\Delta^2-\omega^2}}
\right] 
\left(
\omega - \Delta \cos \frac{\varphi}{2}
\right),
\label{hybrid_WB}
\end{eqnarray}
where $\Theta$ is the Heaviside step function.
The imaginary part of Eq.~(\ref{hybrid_WB}) is traditionally referred to as the hybridization function, see also Sec.~\ref{subsec_nrg}, while the real part is connected to the imaginary one via the Kramers-Kronig relations. 

Taking together, the unitary transformation $\mathbb{T}$ allows to map the original non-diagonal model expressed via $D$ spinor onto a model described in terms of Bogolyubov-type quasiparticles $w_{\sigma}$ coupled to a single normal lead which has an altered TDOS due to the frequency-dependent self-energy $\Sigma^W(\omega^+)$. Moreover, the Bogolyubov quasiparticles $w_{\sigma}$ interact locally via the ordinary Hubbard interaction term. This allows us to redefine the three-terminal setup as one-channel-lead problem similar to the ordinary SIAM and apply NRG in a straightforward way as described in Ref.~\cite{Bulla-1994}. This way, all spectral properties in the spinor basis $W$ can be obtained.

In the Nambu formalism of spinors $D$ spin symmetry is manifestly present and the corresponding Nambu Green function has thus a normal component $G^D_{n}(\omega^+)$ and an anomalous $G^D_{a}(\omega^+)$. Because of Eqs.~(\ref{Hub_w}) and (\ref{G_0w}), spin symmetry is also preserved in the $w$ basis and the resulting Green functions are thus spin independent, i. e. $G^W_{\uparrow }(\omega^+)=G^W_{\downarrow}(\omega^+) \equiv G^W_{n}(\omega^+)$. Even-though they can be directly calculated by means of NRG, in the end, we need to transform back to the original basis of the $d$ electrons. Therefore, one needs to relate the Green functions and their corresponding spectral functions between both bases. Since the unitary transformation $\mathbb{T}$ mixes the original $d_{\sigma}$ and $d_{\sigma}^{\dagger}$ fields only in a linear way, we obtain
\begin{eqnarray}
G^D_{n}(\omega^+)
&=&
\phantom{-}
\frac{1}{2}
\left[
G^W_{n}(\omega^+)
-
G^W_{n}(-\omega^+)
\right],
\\
G^D_{a}(\omega^+)
&=&
-
\frac{1}{2}
\left[
G^W_{n}(\omega^+)
+
G^W_{n}(-\omega^+)
\right],
\end{eqnarray}
where $-\omega^+ = -\omega - i\eta$. Recall that $G^W_{n}(\omega^+)=G^{W*}_{n}(\omega^-)$ with $\omega^- \equiv \omega - i\eta$  and the imaginary part of the Green function equals the spectral function up to a multiplicative factor $-1/\pi$. Therefore, we can directly construct the normal spectral function $A^D_{n}(\omega)$ and the anomalous spectral function $A^D_{a}(\omega)$ in the $d$ basis using
\begin{eqnarray}
A^D_{n}(\omega)
&=&
\phantom{-}
\frac{1}{2}
\left[
A^W_{n}(\omega)
+
A^W_{n}(-\omega)
\right],
\label{symmetrizing}
\\
A^D_{a}(\omega)
&=&
-\frac{1}{2}
\left[
A^W_{n}(\omega)
-
A^W_{n}(-\omega)
\right]
\label{antisymmetrizing}
\end{eqnarray}
with $A_n^W$ being the spectral function corresponding to $G_n^W$. We will often refer to the backwards transformations of the normal spectral and anomalous function as symmetrization and antisymmetrization, respectively.

\section{$\Delta \rightarrow \infty$ case \label{sec_3}}

To demonstrate and assess the concepts derived in Sec.~\ref{sec_2}, we turn first to the well understood $\Delta \rightarrow \infty$ case of the present model and compare it to the standard NRG approach used to solve this limit in Refs.~\cite{Domanski-2017, Tanaka-2007, Oguri-2012, Oguri-2013}. Here, one first applies a $\mathbb{T}$-like transformation to the dot electrons and a combination of $\mathbb{T}$-like and particle-hole transformations to the Wilson chain. As unnoticed by the authors of \cite{Tanaka-2007, Oguri-2012, Oguri-2013}, in the particle-hole symmetric case only the $\mathbb{T}$ transformation to the dot electrons is essential and the rest is just method specific. To show this, we briefly review the approach in Refs.~\cite{Tanaka-2007, Oguri-2012, Oguri-2013}. Crucially, the $\Delta \rightarrow \infty$ Hamiltonian simplifies down to
\begin{eqnarray}
H_{\Delta \rightarrow \infty}
&=&
H_{d,0} + H_U + H_N + H_{T,N} 
\nonumber
\\
&-&
\Delta_{d} 
\left(
\varphi
\right)
\left(
d^{\dagger}_{\uparrow}
d^{\dagger}_{\downarrow}
+
d^{\vphantom{\dagger}}_{\downarrow}
d^{\vphantom{\dagger}}_{ \uparrow}
\right),
\label{DeltaInfty}
\end{eqnarray}
where $\Delta_{d}(\varphi)\equiv\Gamma_S\cos(\varphi/2)$, $H_N$, $H_{d,0}$, $H_{T,N}$ follow our previous notations and $\varepsilon_d$ was considered originally as arbitrary. $H_{\Delta \rightarrow \infty}$ has thus a one-channel-lead form. The BCS effects are present via non-zero off-diagonal terms.

\begin{figure*}
	\includegraphics[width=2.00\columnwidth]{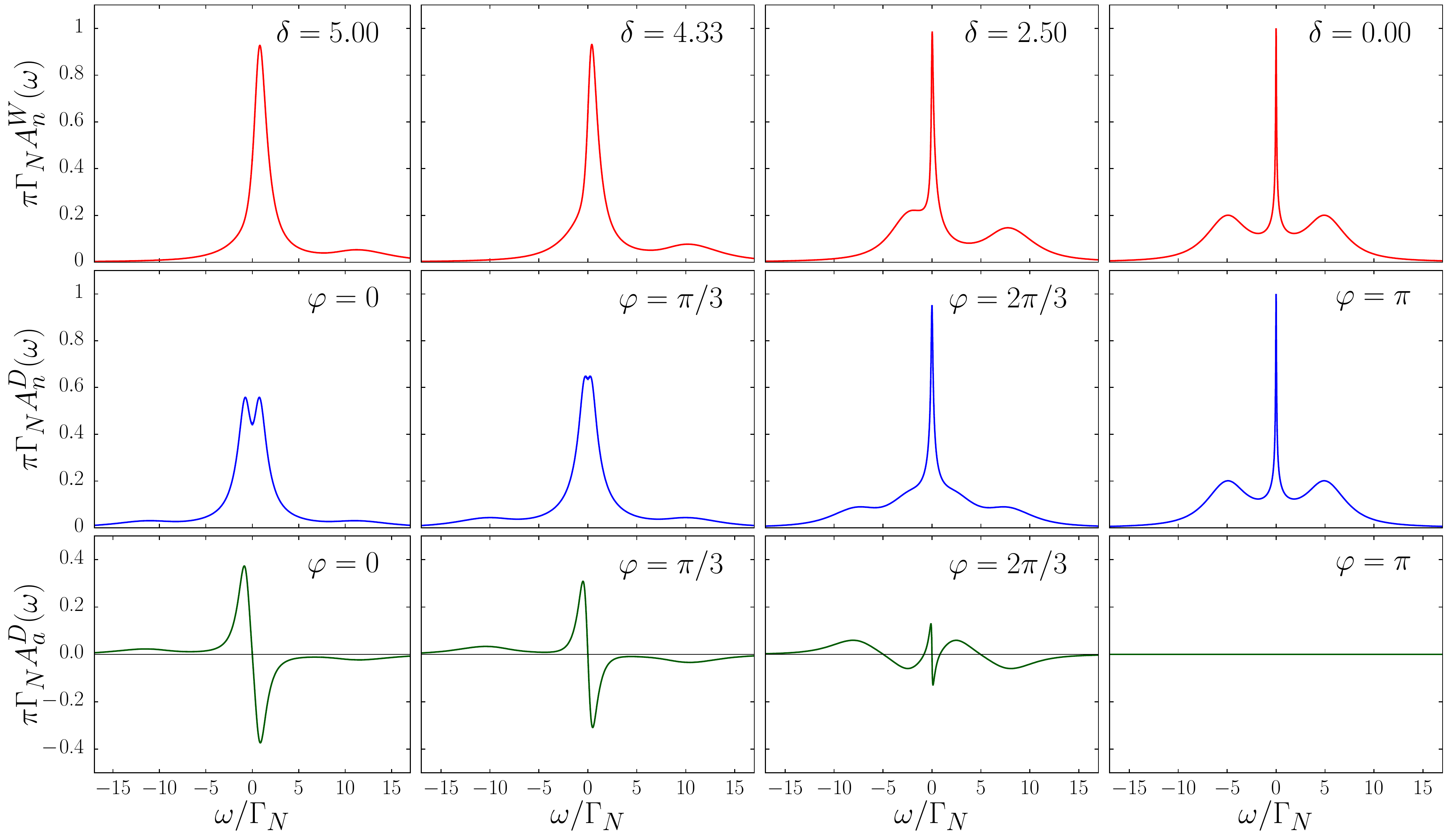}
	\vspace{-0.3cm}
	\caption{
		The top row of panels shows the normal spectral functions for the asymmetric SIAM with $U=5\Gamma_N$ in the direction of decreasing asymmetry parameter $\delta$. The top row spectral functions also correspond to the spectral functions $A^W_n(\omega)$ of the $\Delta \rightarrow \infty$ model expressed in the $w$ basis when $U$ and $\Gamma_N$ are the same and $\delta=\Gamma_s\cos(\varphi/2)$. 
		The middle row of panels shows the normal spectral functions $A^D_n(\omega)$ of the $\Delta \rightarrow \infty$ model in the $d$ basis obtained by symmetrization (\ref{symmetrizing}) of $A^W_n(\omega)$ corresponding to the top row.
		The bottom row shows the anomalous spectral function $A^D_a(\omega)$ of the $\Delta \rightarrow \infty$ model in the $d$ basis obtained by antisymmetrization (\ref{antisymmetrizing}) of the top row.
		\label{fig_2}
	}
\end{figure*}

To treat those, the logarithmically discretized version of $\Delta \rightarrow \infty$ model is mapped onto a semi-infinite Wilson hopping chain with the first node populated by the local $d_{\sigma}$ electrons ($\sigma \in \{\uparrow, \downarrow \}$) of the QD while remaining sites labeled by $i \in N$ are populated by fermions $c_{i \sigma}$ representing the bath degrees of freedom. In Refs.~\cite{Tanaka-2007, Oguri-2012, Oguri-2013}, each site of the Wilson chain is rotated using unitary transformation $O_{\Theta}$
\begin{equation}
O_{\Theta} 
=
\left(
\begin{matrix}
\cos \frac{\Theta}{2} & -\sin \frac{\Theta}{2}
\\[0.4em]
\sin \frac{\Theta}{2} & \phantom{-}\cos \frac{\Theta}{2}
\end{matrix}
\right),
\end{equation}
which acts on the Nambu spinors of the given site of the Wilson chain, while 
\begin{eqnarray}
\cos \frac{\Theta}{2} 
=
\sqrt{\frac{1}{2} + \frac{\tilde{\varepsilon}}{2\delta}},
\qquad
\sin \frac{\Theta}{2} 
=
\sqrt{\frac{1}{2} - \frac{\tilde{\varepsilon}}{2\delta}},
\label{Oangles}
\end{eqnarray}
with $\delta = \sqrt{\tilde{\varepsilon}_d^2 + \Delta^2_d}$ and $\tilde{\varepsilon}_d=\varepsilon_d+U/2$. The first site of the Wilson chain transforms for example as
\begin{equation}
O_{\Theta} D 
=
O_{\Theta}  
\left( 
\begin{matrix}
d^{\vphantom{\dagger}}_{\uparrow}
\vspace{1mm}
\\
d^{\dagger}_{\downarrow}
\end{matrix}
\right)
=
\left( 
\begin{matrix}
w^{\vphantom{\dagger}}_{\uparrow}
\\
w^{\dagger}_{\downarrow}
\end{matrix}
\right)
\equiv
W,
\label{Otransform}
\end{equation}
where $D, W$ follow the notation of Sec.~\ref{sec_2}. The sites representing the lead electrons (index $i$) are also subjected to particle-hole transformations. The coefficients of the resulting diagonal semi-infinite Wilson chain are then noticed to be identical with those of the ordinary asymmetric SIAM and because of the Hausholder transformation the equivalence of the $\Delta \rightarrow \infty$ model to the asymmetric SIAM with the particle-hole asymmetry factor $\delta$ from (\ref{Oangles}) and constant TDOS is established \cite{Tanaka-2007,Oguri-2012, Oguri-2013}. We stress that the findings in Refs.~\cite{Tanaka-2007,Oguri-2012, Oguri-2013} hold at arbitrary filling.

Using the approach of Sec.~\ref{sec_2} we may now prove that in the half-filling, only the unitary transformation $O_{\Theta}$ applied to the space of local electrons is essential while all of the remaining transformations are merely an NRG related technical tool. First, in the half-filled case $\tilde{\varepsilon}_d=\varepsilon_d+U/2=0$ and corresponds to $\Theta=\pi/2$ for which the $O_{\Theta}$ transformation becomes $\mathbb{T}_2^+$ of Sec.~\ref{sec_2} while the self-energy contribution of the leads is known to be proportional to the $2\times2$ unit matrix in the $d$ basis of the local electrons. Thus, it is form invariant under any unitary transformation and $\mathbb{\Sigma}_N^D(\omega)= 
\mathbb{\Sigma}_N^W(\omega)$. The off-diagonal parts of the Hamiltonian (\ref{DeltaInfty}) turn out to be more delicate. Performing then the same partitioning of (\ref{DeltaInfty}) as in Eqs.~(\ref{NambuHdot}) and (\ref{InteractionD}), we may apply the transformation $O_{\pi/2}$ to obtain the non-interacting part of the dot Hamiltonian in a diagonal form
\begin{align}
H_{d,0}
=
W^{\dagger}
\left(
\begin{matrix}
\delta -\frac{U}{2} & 0
\\
0 & -\delta  +\frac{U}{2}
\end{matrix}
\right)
W.
\label{42}
\end{align}
Applying then the transformation $O_{\pi/2}$ to the interaction term gives back (\ref{Hub_w}) which finally proofs that the $\Delta \rightarrow \infty$ model maps onto the asymmetric SIAM with constant TDOS with the particle-hole asymmetry parameter $\delta = \Gamma_S \cos (\varphi/2)$. The resulting Wilson chain is identical to that of Refs.~\cite{Tanaka-2007, Oguri-2012, Oguri-2013}. The physical interpretations of the hybrid reservoir behavior are then easily accessible via the well known results on asymmetric SIAM. Qualitatively, starting from the spectral function $A^W_n(\omega)$ of the asymmetric SIAM of given asymmetry parameter $\delta=\Gamma_S\cos(\varphi/2)$ one applies the symmetrization procedure (\ref{symmetrizing}) and obtains the normal spectral function $A^D_n(\omega)$ of the three-terminal $\Delta \rightarrow \infty$ model in the original basis of the $d$ fields. For the anomalous functions (subscript $a$), analogically, the  antisymmetrization (\ref{antisymmetrizing}) is performed as shown Fig.~\ref{fig_2}. 

In the top row of panels we show the spectral functions of the asymmetric SIAM in descending order of $\delta$ \footnote{The spectral functions have been obtained using the open source NRG Ljubljana code \cite{Ljubljana-code} in the one-channel mode with intertwined $z$-discretization \cite{ZitkoPruschke-2009} with $z= n/10$, $n \in \{ 0, \ldots 10 \}$}. The middle row of panels shows then the symmetrized counterparts of the top row panels which actually represent the solution to the normal spectral functions of the $\Delta \rightarrow \infty$ model at given $\varphi$. Analogically, the bottom row shows the antisymmetrization of the top row panels which then represent the anomalous spectral functions of the $\Delta \rightarrow \infty$ model. Panels in the same columns thus correspond to each other via the relation $\delta = \Gamma_S \cos (\varphi/2)$ and are ordered from left to right with the increasing  phase difference $\varphi$ for the $\Delta \rightarrow \infty$ model and in decreasing order of the asymmetry parameter $\delta$ of the underlying SIAM. The particle-hole symmetric case of the effective SIAM is then realized at $\varphi=\pi$. 

Thus, in the $w$ basis, the particle-hole symmetry at $\varphi=\pi$ leads to the appearance of an ordinary Kondo resonance at the Fermi energy of $A^W_n(\omega)$. Additionally, two satellite Hubbard peaks emerge at approximately $\pm U/2$ (see the last column of panels in Fig.~\ref{fig_2}). Symmetrization (\ref{symmetrizing}) does not alter the shape of the normal spectral function which remains the same in both bases. %In other words, in the $\varphi = \pi$ case, Josephson and other superconducting effects do not disrupt Kondo correlations, which is a trivial result already visible from the effective Hamiltonian (\ref{DeltaInfty}).  

Decreasing the phase difference $\varphi$ and keeping parameters $U$, $\Delta$, $\Gamma_N$ and $\Gamma_S$ constant drives the underlying asymmetric SIAM away from its particle-hole symmetric point as shown in the top row of panels of Fig.~\ref{fig_2} ($\delta$ increases from right to left). As $\delta$ increases, the central Kondo peak shifts gradually away form the Fermi energy and becomes simultaneously broader and slightly smaller \footnote{The broadening has been  also predicted analytically in Refs.~\cite{Domanski-2016,Domanski-2017} via the Schrieffer-Wolff transformation. Using the transformation $\mathbb{T}$, the $\Delta \rightarrow \infty$ can be mapped onto the particle-hole asymmetric SIAM, from which the enhancement of the exchange coupling $J$ compared to the particle-hole symmetric case follows trivially. Thus, as a consequence of the locally induced SC pairing, the $T_K$ is enhanced and the central peak is broader compared to the single-channel normal SIAM.}. When the particle-hole asymmetry in the $w$ basis is relatively small, i. e. $\delta=2.5$ in Fig.~\ref{fig_2}, the off-central movement does not overcome its broadening. Consequently, performing symmetrization operation (\ref{symmetrizing}) to obtain $A_n^D(\omega)$ makes the central peak broader but still singly-peaked. However, decreasing the angle $\varphi$ further eventually causes such a strong decentralization that the broadening is insufficient and symmetrization (\ref{symmetrizing}) then only leads to a split central peak with the remnants of Kondo resonances, as seen at larger $\varphi$ corresponding to $\delta=4.33$ and $\delta=5.0$ cases in Fig.~\ref{fig_2}. At such a critical value, $\varphi^*\approx \pi/3$ in Fig.~\ref{fig_2}, the splitting of $A^D_n(\omega)$ is related to sufficient suppression of the Kondo correlations in the $w$ basis. However, keeping the value of $\Gamma_S$ so small that for given interaction strength $U$ the asymmetry parameter $\delta$ is insufficient to destroy Kondo correlations, splitting might be avoided in analogy to SCIAM.

At $\varphi=0$ (the first column of Fig.~\ref{fig_4}), the split-peak is accompanied by a highly suppressed pair of peaks at $\omega \approx \pm 12 \Gamma_N$ which intensifies and shifts towards the Fermi energy as the angle $\varphi$ is increased because the corresponding charge excitations of the underlying asymmetric SIAM become stronger ($\delta$ decreases). Therefore, at $\varphi^* \approx \pi/3$, when the split-peak merges into a Kondo-like central peak, this pair becomes well visible. Moreover, a second pair starts to emerge from the Kondo-like peak (note the shoulders of the central peak at $\varphi=2\pi/3$ in Fig.~\ref{fig_2}). These two pairs move then towards $\pm U/2$ until they merge at $\varphi=\pi$ where they correspond to the ordinary Hubbard peaks of the symmetric SIAM. The behavior of the off-center peaks thus highly resembles that of the ABS states in the SCIAM (apart of the existence of two pairs also for $\varphi<\varphi^*$).

%Altogether, the spectral function $A^D_n(\omega)$ at $\varphi < \varphi^*$ has a pair of symmetrically placed intensive peaks while the outer pair of peaks is mostly unnoticeable. At $\varphi \approx \varphi^*$ the inner peaks cross each other and continue to move towards $\pm U/2$ with increasing $\varphi$. The pair of the outer peaks becomes more pronounced at around $\varphi^*$ while simultaneously shifting towards the Fermi energy until $\varphi=\pi$, where they finally merge together with the inner pair of peaks at approximately $\pm U/2$. Such 

The normal lead of the $\Delta \rightarrow \infty$ model causes not only a singlet ground state for all values of $\varphi$ but also lifts the strict selection rules present in SCIAM which explains the additional pair of peaks for $\varphi<\varphi^*$. Thus, for $\varphi<\varphi^*$ the spectral function $A^D_n(\omega)$ resembles somewhat broadened spectral function of the $0$ phase of SCIAM and is thus referred to as $0$-like phase in what follows. For $\varphi>\varphi^*$, $A^D_n(\omega)$ obtains a shape similar to that of broadened spectral function in the $\pi$ phase of SCIAM. However, due to the normal electrode a Kondo resonance coexists with four broadened ABS states.

\section{Finite-gap model \label{sec_4}}

\begin{figure}[t]
	\begin{center}
		\vspace{-1cm}
		\includegraphics[width=1.00\columnwidth]{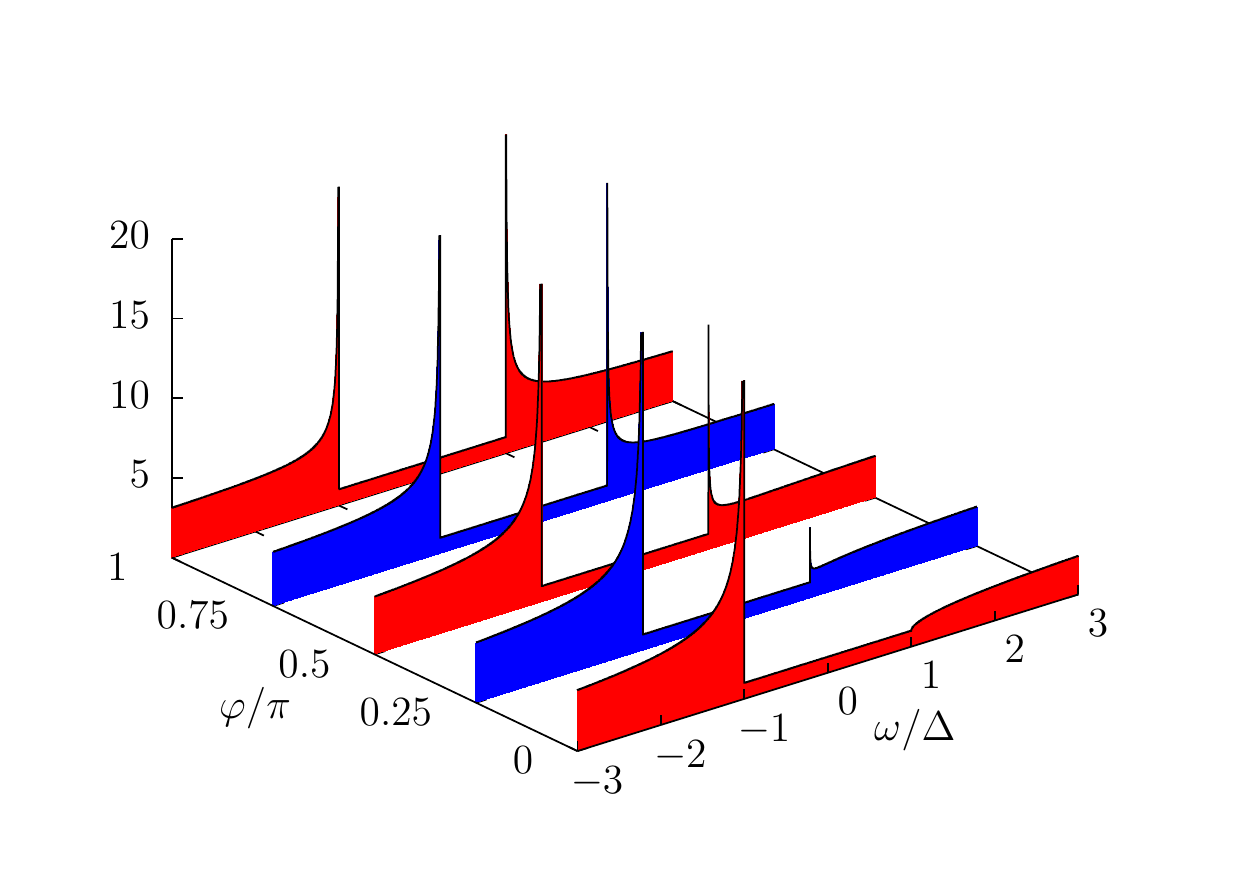}
		\vspace{-1cm}
		\caption{Phase evolution of the TDOS $\Gamma^W(\omega)$ \eqref{TDOS_WB} in the $w$ basis for $\Gamma_N=\Delta$ and $\Gamma_S=2\Delta$. At $\varphi = 0$, only the left BCS singularity does appear and the resulting TDOS is highly asymmetric. Increasing $\varphi$ diminishes the asymmetry while BCS singularities develop at both gap edges. The symmetry is fully restored only at $\varphi=\pi$.  \label{fig_3}   }
	\end{center}
\end{figure}

\subsection{NRG calculations \label{subsec_nrg}}

As shown in Sec.~\ref{subsec_transform}, the transformation $\mathbb{T}$ maps the finite-gap three-terminal setup at the half-filling onto an NRG-tractable one-channel problem. In the $w$ basis, the Hamiltonian describes an Anderson impurity coupled to a continuum of bath states with modified TDOS $\Gamma^W(\omega)$ corresponding to $\Sigma^W(\omega)$ which except of $\varphi = \pi$ is particle-hole asymmetric as shown in Appendix~\ref{app_finite}. In the limit of infinitely wide band it reads
\begin{equation}\label{TDOS_WB}
\Gamma^W(\omega)=\Gamma_N+\frac{\Gamma_S|\omega|\Theta(\omega^2-\Delta^2)}
{\sqrt{\omega^2-\Delta^2}}\left(1-\frac{\Delta }{\omega}\cos\frac{\varphi}{2}\right),
\end{equation}
where $\Theta$ is the Heaviside step function. The phase evolution of hybridization function $\Gamma^W(\omega)$ is shown in Fig.~\ref{fig_3} for selected parameters. Note, that $\Gamma^W(\omega)$ is only particle-hole symmetric at $\varphi = \pi$ with asymmetry increasing towards $\varphi=0$ in analogy to the $\Delta \rightarrow \infty$ case. Since $\Gamma^W(\omega)$ is diagonal, in the $w$ basis standard one-channel NRG method of Refs.~\cite{Bulla-1994, Bulla-Rev-2008} can be applied. To this end, we have utilized NRG Ljubljana code \cite{Ljubljana-code} with intertwined $z$-discretization according to the scheme of \v{Z}itko et al. \cite{ZitkoPruschke-2009}, i. e. $z= n/10$ where $n \in \{ 0, \ldots 10 \}$. To achieve smoother spectral functions with discontinuities at the BCS gap edges, the so-called self-energy trick has been employed. We stress that in the main body of the article we concentrate on the wide band limit with bandwidth set to $2B=4000\Delta$. The corrections for the case of a narrow band are discussed in the Appendix~\ref{app_finite}.

The experimentally accessible spectral functions in the $d$ basis have been obtained by means of Eqs.~(\ref{symmetrizing}) and (\ref{antisymmetrizing}) and the results are discussed in Sec.~\ref{subsec_spec}. On-dot induced pairing $\nu=\langle d_{\downarrow}d_{\uparrow} \rangle $ is trivially connected to the filling $n_w$ in the $w$ basis and can be measured directly as discussed in Sec.~\ref{subsec_transp}. The operator for the Josephson current depends explicitly from the lead electrons and an integral formula of Ref.~\cite{Zonda-2015} involving the anomalous component of Green function in the $d$ basis is required as discussed in Sec.~\ref{subsec_transp}.

\begin{figure}[t]
	\includegraphics[width=1.00\columnwidth]{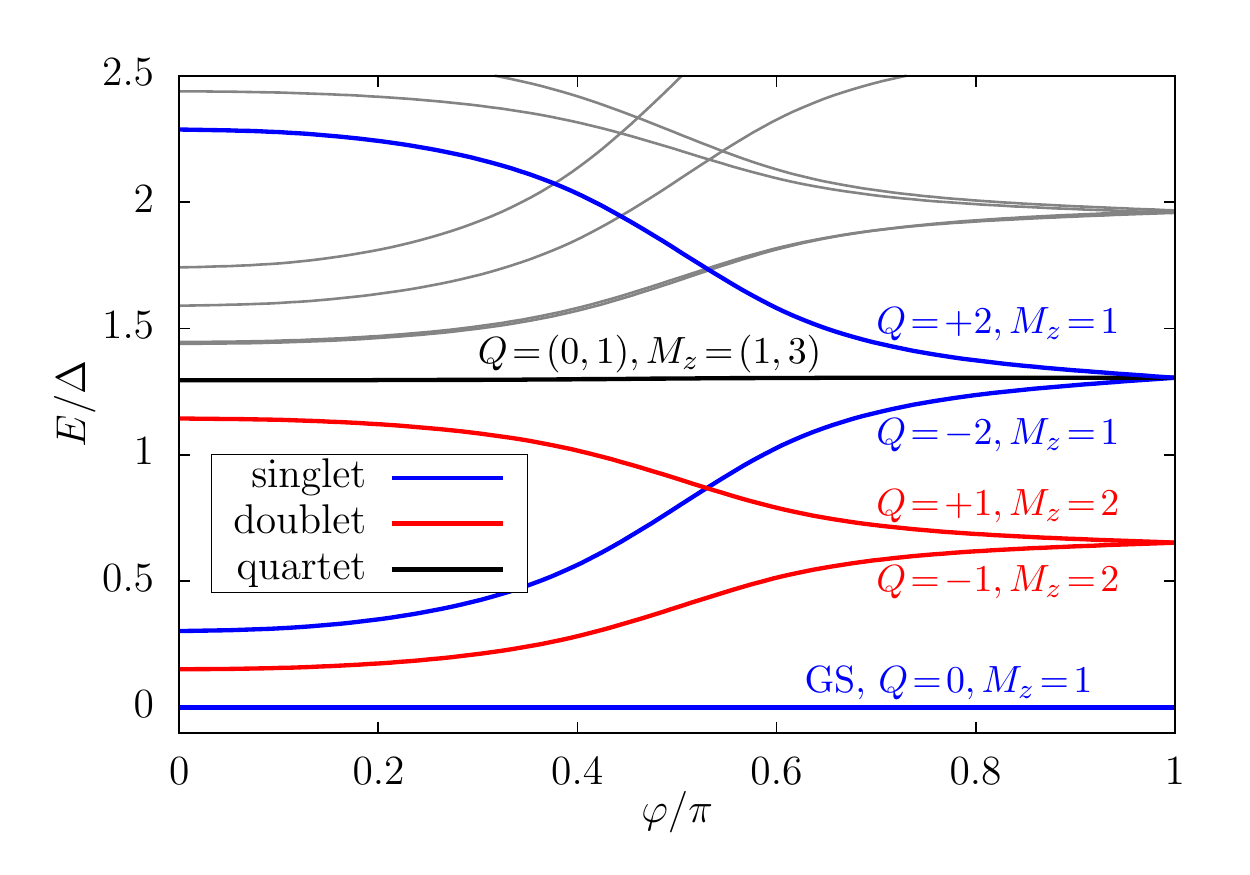}
	\caption{Low energy eigenvalues of the logarithmically discretized Hamiltonian obtained using NRG Ljubljana for the half-bandwidth $B=2000\Delta$, $U=3\Delta$, $\Gamma_S=\Delta$ and $\Gamma_N=\Delta/100$. $Q$ is defined as the total charge of the Wilson chain measured with respect to the half-filling and $M_z \equiv 1 + 2S_z$ with $S_z$ being the overall magnetization of the Wilson chain.  \label{fig_4}  }
\end{figure}

Most importantly, we note that the energy eigenvalues, as obtained at each NRG iteration, are basis independent because $\mathbb{T}$ is unitary. Corresponding effective models attributed to certain RG fixed point can thus be directly read off. Here, we concentrate exclusively at low temperature behavior which is governed by the strongly coupled (SC) or frozen impurity (FI) RG fixed point depending on the extent of phase-bias-induced particle-hole asymmetry in the $w$ basis as shown in Fig.~\ref{fig_4}. Here, we selected parameters involving a sign reversal of the local pairing at $\varphi ^*_{\mathrm{pair}}\approx 0.45 \pi$ and Josephson current reversal at $\varphi^*_j \approx 0.5 \pi$ (see the discussion in Sec.~\ref{subsec_transp}). At $\varphi=\pi$, the SC fixed point of ordinary symmetric SIAM is identified since it contains a singlet ground state, followed by first a quadruplet and then a sextet of next excited levels. Decreasing $\varphi$, splits the quadruplet into two doublets while the sextet splits into two singlets placed symmetrically around the remaining quadruplet corresponding to the behavior of ordinary asymmetric SIAM at small particle-hole asymmetry. This, establishes then a correspondence of the present three-terminal set-up to the particle-hole asymmetric SIAM at $\varphi \neq \pi$ with particle-hole symmetric case recovered at $\varphi=\pi$. The qualitative behavior of the spectral functions is therefore expected to essentially follow the results of the $\Delta \rightarrow \infty$ case discussed in Sec.~\ref{sec_3}. The quantitative changes in on-dot induced pairing and Josephson current are therefore only related to the fine details of the corresponding spectral functions at higher frequencies as discussed in Secs.~\ref{subsec_spec} and \ref{subsec_transp}.

\subsection{Spectral properties and the Kondo scale \label{subsec_spec} }

\begin{figure*}
	\begin{center}
		\includegraphics[width=2.0\columnwidth]{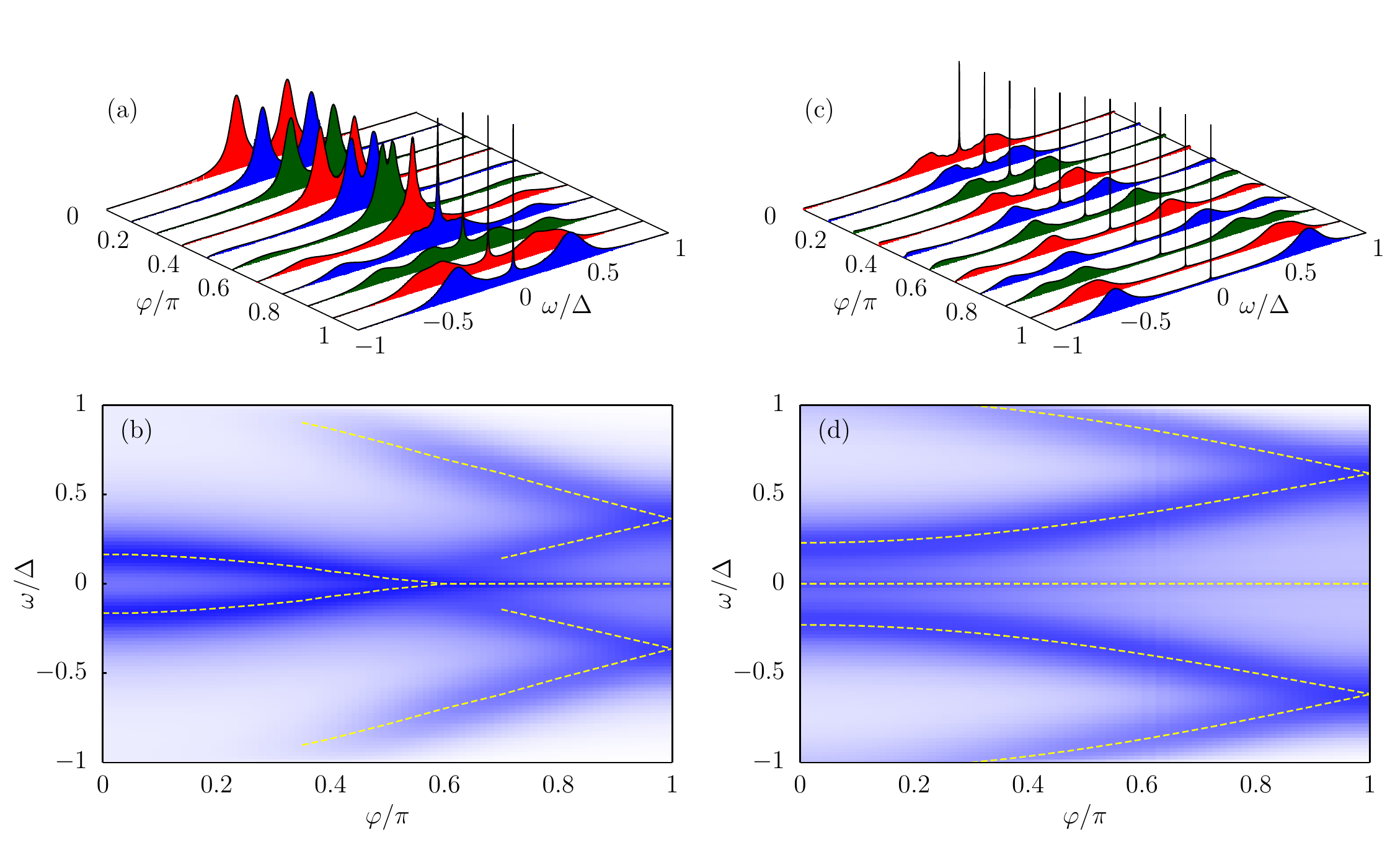}
		\caption{
		$(a)$ 
		Phase evolution of the spectral function $A_n^D(\omega)$ in the sub-gap region of the finite-gap three-terminal setup for $U=3\Delta$, $\Gamma_N=\Delta/10$, $\Gamma_S=\Delta$ and half-bandwidth $B=2000\Delta$. For $\varphi<\varphi^*\approx 0.55\pi$ one observes a pair of broadened ABS states while for $\varphi>\varphi^*$ an additional pair of broadened ABS states and an additional central Kondo-like peak do appear. In an analogy to the SCIAM, for $\varphi<\varphi^*$ the regime is referred to as the $0$-like phase while for $\varphi>\varphi^*$ as the $\pi$-like phase. 
		$(b)$ 
		Heatmap corresponding to panel $(a)$ highlights the phase-dependent position of the maxima of the in-gap peaks (dashed lines). 
		$(c)$
		The same as in panel $(a)$ only $U=6\Delta$. The Kondo correlations dominate the system which remains in the $\pi$-like phase for all values of $\varphi$. For sufficiently large $U$, such a scenario does occur also in the SCIAM.
		$(d)$ 
		Heatmap corresponding to panel $(c)$ shows the phase-dependent position of the maxima of the in-gap peaks (dashed lines). 
		\label{fig_5} }
	\end{center}
\end{figure*}

The phase evolution of the normal spectral function in the $d$ basis shown in Fig.~\ref{fig_5} for two values of $U/\Delta$ demonstrates that qualitatively finite-gap case does not differ much from the $\Delta \rightarrow \infty$ case. Selecting first the $U=3\Delta$ case (panels $(a)$ and $(b)$ in Fig.~\ref{fig_5}), we notice two broadened ABS-like peaks placed symmetrically around the Fermi energy at $\varphi=0$. With increasing $\varphi$, both peaks move towards the Fermi energy and merge at a certain value $\varphi^*$ which depends non-trivially on $U$ and $\Delta$. Subsequently, for all $\varphi>\varphi^*$ the central Kondo-like peak is present. Moreover, four side-peaks (corresponding to the two symmetrized Hubbard satellites in the $w$ basis) also emerge. Increasing $\varphi$ further shifts the two peaks on each side of the spectra together, until at $\varphi=\pi$ they coalesce at approximately $\pm U/2$. At this point, the TDOS is symmetric and the $\varphi=\pi$ spectrum resembles the typical three-peak structure of the symmetric SIAM. The second case with $U=6\Delta$ is shown in panels $(c)$ and $(d)$ of Fig.~\ref{fig_5}. Here, the ratio $\Gamma_S/\Delta$ is insufficient to generate particle-hole asymmetry leading to the emergence of the $0$-like phase. Such regimes are analogous to the observations made for SCIAM at large ratios $U/\Delta$.

To make the movement of the in-gap peaks explicitly manifest, we visualized the phase-dependent positions of their maxima via heatmaps in Fig.~\ref{fig_5}, panels $(b)$ and $(d)$. We clearly observe in panel $(b)$ of Fig.~\ref{fig_5} their crossing at angle $\varphi^*$. When $\varphi$ is further increased, the in-gap peaks move apart again. However, for all $\varphi>\varphi^*$ two additional in-gap states emerge relatively close to the band edges. The two peaks for $\varphi<\varphi^*$ can thus be related to the two ABS states of SCIAM in the $0$ phase, while the four off-central peaks are in one-to-one correspondence with the four ABS states observed in the $\pi$ phase of SCIAM. However, unlike in SCIAM the non-zero TDOS around the Fermi energy gives rise also to the central Kondo-like resonance for all $\varphi>\varphi^*$.

Thus, the obtained spectral functions qualitatively correspond to the $\Delta \rightarrow \infty$ model and the physical interpretation in terms of the particle-hole asymmetry of the underlying model in the $w$ basis holds analogously. Nevertheless, there are quantitative differences which appear once integral quantities, such as the filling $n_w$ in the $w$ basis, are considered. For the $\Delta \rightarrow \infty$ model,  the filling $n_w$ monotonically decreases from the $n_w=1$ value obtained at $\varphi=\pi$ for all parameter regimes. In the finite-gap three-terminal case, there are parameter regimes where $n_w$ first increases to values larger than $1$ (positive effective chemical potential) and then starts to monotonically decrease to values smaller than $1$ (negative effective chemical potential). Such integral properties are  shown in Sec.~\ref{subsec_transp} to be crucial for the system to exhibit effects such as pairing or Josephson current reversal which are typical of $0$-$\pi$ transition observed in SCIAM. This means that the precise shape of spectral functions plays an important role when analyzing the finite-gap case.  

%Let us now describe the behavior in the $0$-like phase in more detail. The pair of strongly pronounced in-gap peaks for $\varphi<\varphi^*$ can easily be interpreted as the broadened ABS states of the purely superconducting case. To this end, it can be verified that their structure can be fitted well by Lorentzian line shapes centered around the frequency $\omega_{\mathrm{ABS}}$ of the broadened ABS maxima with $\Gamma_N$ used as the broadening parameter. Consequently, for $\varphi<\varphi^*$ the spectra may be not only qualitatively but also quantitatively interpreted as broadened ABS states of the pure SCIAM model. However, once $\varphi$ approaches $\varphi^*$ the broadening of the in-gap states becomes much larger and for $\varphi>\varphi^*$ it does not match with $\Gamma_N$. On the other hand, the number of the in-gap peaks is still four and they correspond well to the ABS states of the $\pi$ phase of the SCIAM. 

%Let us now turn to the quantitative analysis of the $\pi$-like phase, where an unusual co-existence of the Kondo-like peak with four broadened ABS states ($\varphi>\varphi^*$) is observed. Since the prerequisite of an effective Kondo screening is the non-zero TDOS around the Fermi energy brought in by the normal electrode, the central peak forms for arbitrary values of $U$ at $\varphi=\pi$. 

In Sec.~\ref{subsec_nrg}, we have already established the correspondence of the low energy many-body NRG spectra to that of the particle-hole asymmetric SIAM for $\varphi \neq \pi$ with particle-hole asymmetry monotonically decreasing towards $\varphi=\pi$ where it completely vanishes. In the $w$ basis, one therefore observes that starting at the particle-hole symmetric case for $\varphi=\pi$ and then decreasing $\varphi$, causes a gradual movement of the original Kondo peak away from the Fermi energy which is induced by the increasing particle-hole asymmetry. However, as long as $\varphi>\varphi^*$ the increasingly large broadening does overcome this shift and symmetrization (\ref{symmetrizing}) still leads to a well defined central peak in the spectral function $A^W_{n}(\omega)$. Only when $\varphi$ is decreased further, does the broadening of the central peak stop compensating for the rapid movement of the peak, so that the symmetrization (\ref{symmetrizing}) results in a doubly peaked spectral function $A^W_{n}(\omega)$ in the $w$ basis. 

\begin{figure}[ht]
	\begin{center}
		\includegraphics[width=1.0\columnwidth]{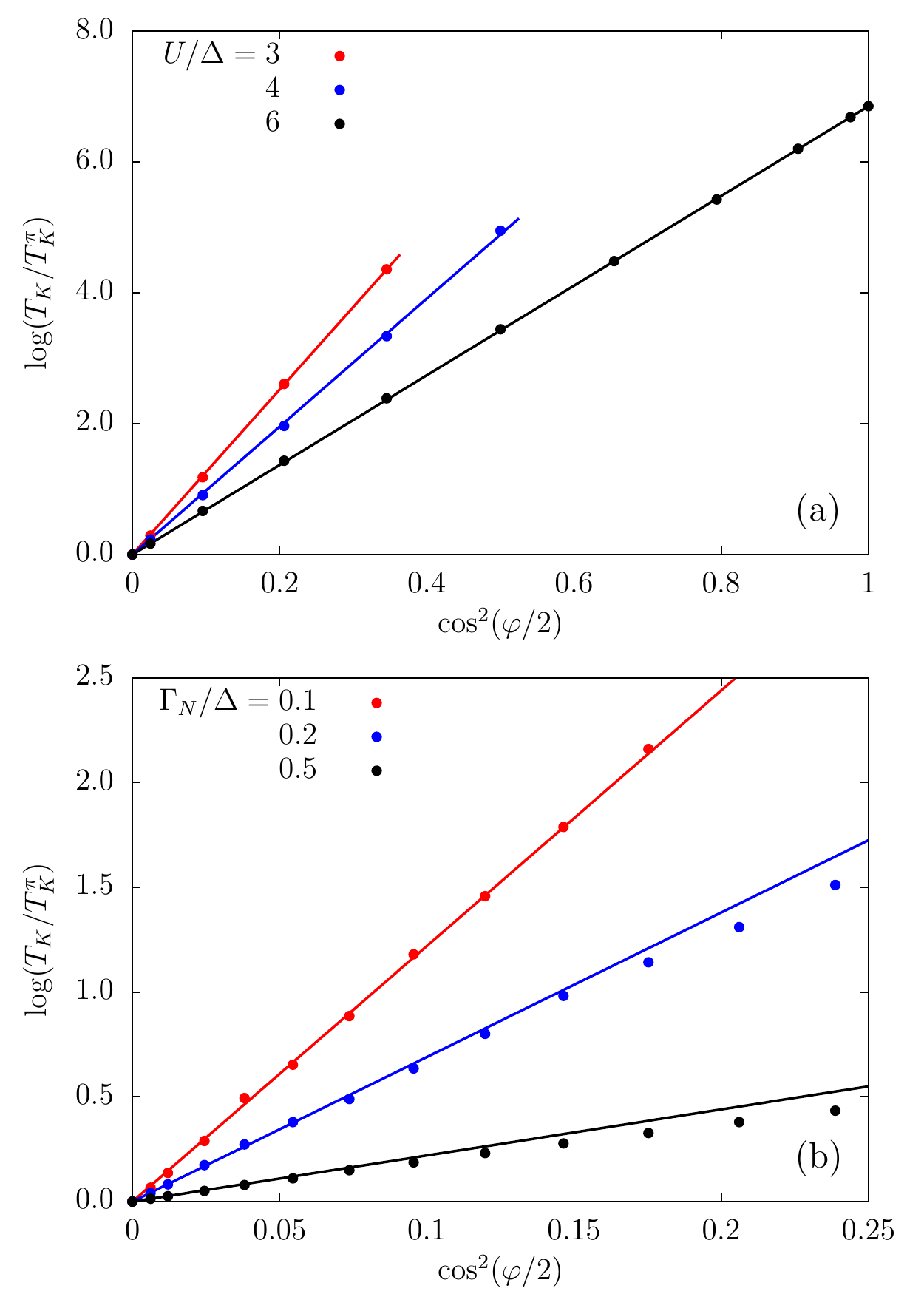}
		\caption{
			$(a)$
			Phase dependence of the Kondo temperature $T_K$ of the three-terminal set-up at $\Gamma_S=\Delta$, $\Gamma_N=\Delta/10$ at varying $U$ with half-bandwidth $B=2000\Delta$. $T_K$ is 
			determined as HWHM of the central Kondo-like peak observed in the $\pi$-like phase, $T_K^{\pi}$ denotes $T_K$ at $\varphi=\pi$. Points represent NRG data while lines are fits in the corresponding $\pi$-like phase regions. In all three cases, $\log T_K \propto \cos^2(\varphi/2)$ and follows the hypothesis of Ref.~\cite{Domanski-2017}.
			$(b)$
			The same as in panel $(a)$ at fixed $U=3\Delta$ and varying $\Gamma_N$. The solid lines now represent tangents at $\cos(\varphi/2)=0$ ($\varphi=\pi$). For higher values of $\Gamma_N$ the $0-\pi$-like crossover region is only insignificantly shifted towards lower values of $\varphi$ (higher values of $\cos(\varphi/2)$). The range of the horizontal axis is thus selected to be narrower compared to panel $(a)$, i.e.: $0 \leq \cos^2(\varphi/2) \leq 1/4$ ($2\pi/3 \lesssim \varphi \leq \pi$). Exceeding $\Gamma_N=\Delta/5$ we observe clear deviations from the hypothesis of Ref.~\cite{Domanski-2017} which are due to the BCS electrons directly entering the formation of the Kondo resonance.
			\label{fig_6}
		}
	\end{center}
\end{figure} 

The lack or presence of a single central peak can thus be understood as a sign of the Kondo-like interaction-screening efficiency. To quantify such behavior, we have extracted the phase-dependent Kondo temperature $T_K$ as the half-width at half maximum (HWHM) value of the zero-energy peak of the $\pi$-like phase. Unlike in Ref.~\cite{Domanski-2017}, we first compare $T_K$ to the Kondo temperature at $\varphi=\pi$, denoted as $T_K^{\pi}$, which as shown by the transformation $\mathbb{T}$ preserve particle-hole symmetry in the $w$ basis. Decreasing $\varphi$ from its particle-hole symmetric point at $\varphi=\pi$ then introduces increasingly larger particle hole asymmetry in the $w$ basis and is also accompanied by the enhancement of $T_K$ in the experimentally relevant $d$ basis. Such a phase-dependent enhancement of $T_K$ is then conveniently measured via $\log(T_K/T_K^{\pi})$ as done in Fig.~\ref{fig_6}.

\begin{figure}[ht]
	\begin{center}
		\includegraphics[width=1.0\columnwidth]{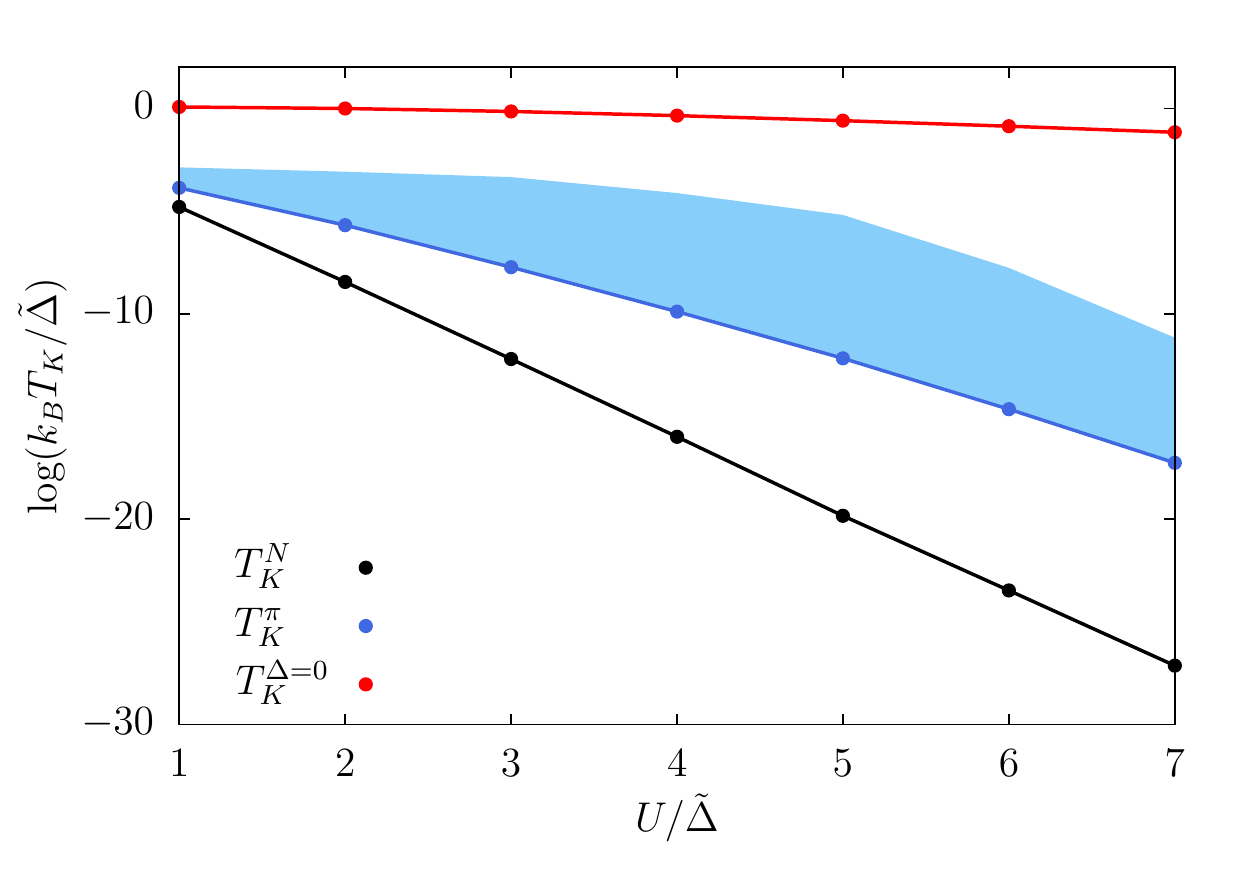}
		\caption{
			Dependence of the Kondo temperature $T_K$ on $U$ for the three-terminal set-up (blue) and analogous systems with decoupled BCS leads (black) and with a closed BCS gap (red). $T_K$ in all systems is defined via HWHM as in Fig.~\ref{fig_6}. Points represent NRG data while lines are just for visual guidance. We use $\tilde{\Delta}=B/2000$ as the unit of energy. The full three-terminal set-up (blue) is calculated at parameters $\Delta=\tilde{\Delta}$, $\Gamma_S=\tilde{\Delta}$ and $\Gamma_N=\tilde{\Delta}/10$. Blue points and the blue solid line represent values at $\varphi=\pi$ while the blue shaded region covers all $\varphi$ values in $\pi$-like phase. Values for the decoupled BCS leads are obtained by setting $\Gamma_S=0$ and keeping $\Gamma_N=\tilde{\Delta}/10$ in the full three-terminal set-up. The case of the closed gap is obtained by setting $\Delta=0$
			in the full three-terminal set-up and keeping the remaining parameters the same. It corresponds to the ordinary SIAM with  the combined hybridization strength of all three metallic leads $\Gamma_N+\Gamma_S=1.1\tilde{\Delta}$. Clearly, the Kondo temperature of the full three-terminal set-up is bounded between the other two cases. 
			\label{fig_7}
		}
	\end{center}
\end{figure}

%Now, let us first consider the qualitative features connected to the phase-dependent enhancement of $T_K$ in the region of the $\pi$-like phase. As follows from the unitary transformation $\mathbb{T}$ derived in Sec.~\ref{sec_2}, the underlying cause of the phenomena is related to the increasing particle-hole asymmetry in the transformed $w$ basis. 
The broadening of the central peak in the experimentally observed $d$ basis, as encoded by $T_K$, is thus accomplished by a delicate interplay between the shift and broadening of the central peak in the $w$ basis due to the increase of the particle-hole asymmetry. Once the symmetrization (\ref{symmetrizing}) is applied, the two effects combine to a wide and somewhat deformed central peak in the $d$ basis as compared to the particle-hole symmetric case $\varphi=\pi$. The increasing $T_K$ therefore cannot be completely attributed to the increase of Kondo correlations, as speculated in Ref.~\cite{Domanski-2017}, as charge fluctuations become more important and may even lead to complete destruction of the central Kondo peak once the crossover to the $0$-like phase is entered. Consequently, the enhancement of $T_K$ with decreasing $\varphi$ is to be attributed to the increase of the underlying particle-hole asymmetry of the system as seen in the $w$ basis introduced in Sec.~\ref{sec_2}.
		
Predictions on the phase-dependent enhancement according to $\log T_K\propto\cos^2(\varphi/2)$ were performed already in Ref.~\cite{Domanski-2017} but were based only on numerical indications from the second order perturbation theory. To thoroughly assess this conjecture we thus use the exact NRG data for $\Gamma_N=\Delta/10$ at various $U$, see the upper panel of Fig.~\ref{fig_6}. The phase-dependent spectral functions corresponding to $U=3\Delta$ and $U=6\Delta$ cases have already been presented in Fig.~\ref{fig_5} from which it is evident that the region of $\pi$-like phase is considerably increasing with $U$ at fixed $\Gamma_N$. Since the HWHM as a measure of $T_K$ becomes meaningless below a given $\varphi^*$, the corresponding dependencies in the upper panel of Fig.~\ref{fig_6} terminate at their corresponding $\varphi^*$ and only the case $U=6\Delta$ covers the whole available $\varphi$ range. Nevertheless, in all three cases we observe that $\log T_K$ is proportional to $\cos^2(\varphi/2)$ with no significant deviations appearing when approaching the crossover region around $\varphi^*$. In these parameter regimes, the hypothesis of Ref.~\cite{Domanski-2017} is thus very well satisfied.

However, when $\Gamma_N$ is increased up to the size comparable with $\Delta$, we expect the divergent portion of $\Gamma^W(\omega)$ present at the gap edges to become more involved in the formation of the Kondo resonance. This in turn potentially deforms the central peak and may cause deviations from the $\log T_K \propto \cos^2(\varphi/2)$ law observed previously. To investigate such regime, we selected the $U=3\Delta$ case shown previously and increased $\Gamma_N$, see panel $(b)$ of Fig.~\ref{fig_6}. Since, the size of the $\pi$-like phase region is almost independent of $\Gamma_N$, we may focus onto the narrower range of $ 2\pi/3 \lesssim \varphi < \pi$ ($0  <  \cos^2(\varphi/2) \leq 1/4$). For $\Gamma_N=\Delta/5$ and $\Gamma_N=\Delta/2$ the phase-dependencies obtained by NRG (points in the graph) follow the tangents at $\varphi=\pi$ (solid lines in panel $b)$ of Fig.~\ref{fig_6}) quite closely. However, they start to deviate increasingly in the crossover region as $\varphi$ is decreased towards $\varphi^*$. The deviations from the hypothesized $T_K \propto \cos^2(\varphi/2)$ law are however unrelated to entering the crossover region of the $\pi$-like to $0$-like transition as follows from the for $\Gamma_N=\Delta/10$ case (red points in panel $b)$ of Fig.~\ref{fig_6}).

Let us now connect the findings to the expected experimental outcome. In the literature, there are numerous statements referring the {\em enhancement} of the Kondo scale upon switching on the superconductivity in the three-terminal setup  \cite{Domanski-2017,Domanski-2016,Fazio-1998}. The issue is, however, which reference system is used for the comparison (how the switch-on of the superconductivity is achieved). One option is to {\em add} the superconducting lead(s) to the conventional Anderson/Kondo model of a QD with one normal lead as used, for example, in Ref.~\cite{Domanski-2016} for the case of one added superconducting lead in the $\Delta\to\infty$ model. The resulting enhancement of $T_K$ due to the addition of the superconducting lead is indisputable, nevertheless, in experiments, it would require the possibility of a controlled tunnel coupling/decoupling of the superconducting lead. While it is in principle possible by electrostatic gating of the pinch-off of the tunneling connection, the conventional experimental practice works differently --- the reference normal system would not consist of the single normal lead but of all involved leads, including the superconducting one(s), turned into the normal state by a small magnetic field.            

To quantify this matters, we first evaluate $T_K$ of the three-terminal set-up with $\Gamma_S=10$, $\Gamma_N=\Delta=B/2000$ ($2B$ being the bandwidth) at various interaction strengths $U$ and for all phase differences compatible with $\pi$-like phase. The resulting values of $T_K$ lie all in the blue shaded region of Fig.~\ref{fig_7} with $\varphi=\pi$ cases being represented by the blue line in Fig.~\ref{fig_7}). First, we consider referencing the outcome against the both BCS electrodes completely decoupled from the present three-terminal set-up. Since $\Gamma_S=0$, we note that the system is just the ordinary particle-hole symmetric SIAM with constant TDOS given by $\Gamma_N$. We thus denote the corresponding Kondo temperature as $T_K^N$ and vary the interaction strength $U$ at constant $\Gamma_N$ (black line in Fig.~\ref{fig_7}). Clearly, $T_K>T_K^N$ for all values of $U$ in the plotted region and further by extrapolation. Consequently, enhancement of the Kondo screening due to the additional BCS correlations is verified in accord with Refs.~\cite{Domanski-2017,Domanski-2016}. The second, experimentally more accessible option is obtained by setting $\Delta=0$ (the phase-dependence vanishes). The corresponding Kondo temperature is denoted then $T_K^{\Delta=0}$ (red line in Fig.~\ref{fig_7}). Now, $T_K<T_K^{\Delta=0}$ for all plotted values of $U$ and further by extrapolation.  The introduction of the superconducting correlations is thus clearly {\em decreasing} $T_K$, which is also our prediction for the conventional experimental setups.

\subsection{Pairing and Josephson current \label{subsec_transp} }

We now briefly address the transport properties in the hybrid three-terminal structure. Because they have already been obtained for finite temperatures in Ref.~\cite{Domanski-2017} using QMC, we will mostly concentrate on the methodology in our present approach and use the available results as a comparison to the $T=0$ results presented here.

Although, the transformation $\mathbb{T}$ allows simpler Hamiltonian formulation of the present problem in the $w$ basis, all transport properties are naturally measured in the original $d$ basis. Superconducting effects are then related to the off-diagonal terms of the Hamiltonian, or equivalently to the off-diagonal Nambu Green functions,  expressed in the $d$ basis. Thus, for example, although the on-dot induced pairing in the $w$ basis is by definition zero as $G^W(\omega)$ has no off-diagonal entries, one may show by simple application of the transformation $\mathbb{T}$ to the definition of $\nu$ that
\begin{equation}
\nu
\equiv
\braket{ d_{\downarrow} d_{\uparrow} }
=
\frac{1}{2}
- 
\frac{n_w}{2},
\label{PairingMap}
\end{equation}
where $n_w\equiv n_{w\uparrow}+n_{w\downarrow}$ is the sum of occupations of the spin-up and spin-down levels in the $w$ basis with $n_w=1$ in half-filling. Interestingly, changing $n_w<1$ to $n_w>1$ in the above equation induces then sign reversal of $\nu$.

Consequently, for $\Delta \rightarrow \infty$ model the mapping onto the ordinary asymmetric SIAM effectively prohibited any sign reversal of $\nu$ as $n_w<1$ strictly. In the case of the finite-gap three-terminal setup, such restrictions are lifted and the observed dependencies resemble closely the behavior of SCIAM (see Fig.~\ref{fig_7}). However, unlike in SCIAM no true phase transition is present and $\nu$ is a continuous function of $\varphi$ with a crossover region of significant drop only visible for small $\Gamma_N/\Gamma_S$. For $\Gamma_N=\Gamma_S/100$ and $\Gamma_N=\Gamma_S/10$, $\nu$ even reverses sign at $\varphi^*_{\mathrm{pair}}$.  However, $\varphi^*_{\mathrm{pair}} \neq \varphi^*$ as the two values coincide only for $\Gamma_N \rightarrow 0$. The comparison of the present results to QMC is shown in the the Appendix~\ref{app_qmc}, where also the effects of the finite bandwidth are discussed. At this place, it is sufficient to note that deep in the $0$-like or $\pi$-like phase QMC and NRG agree well within their numerical accuracy. 

\begin{figure*}[ht]
	\includegraphics[width=2.00\columnwidth]{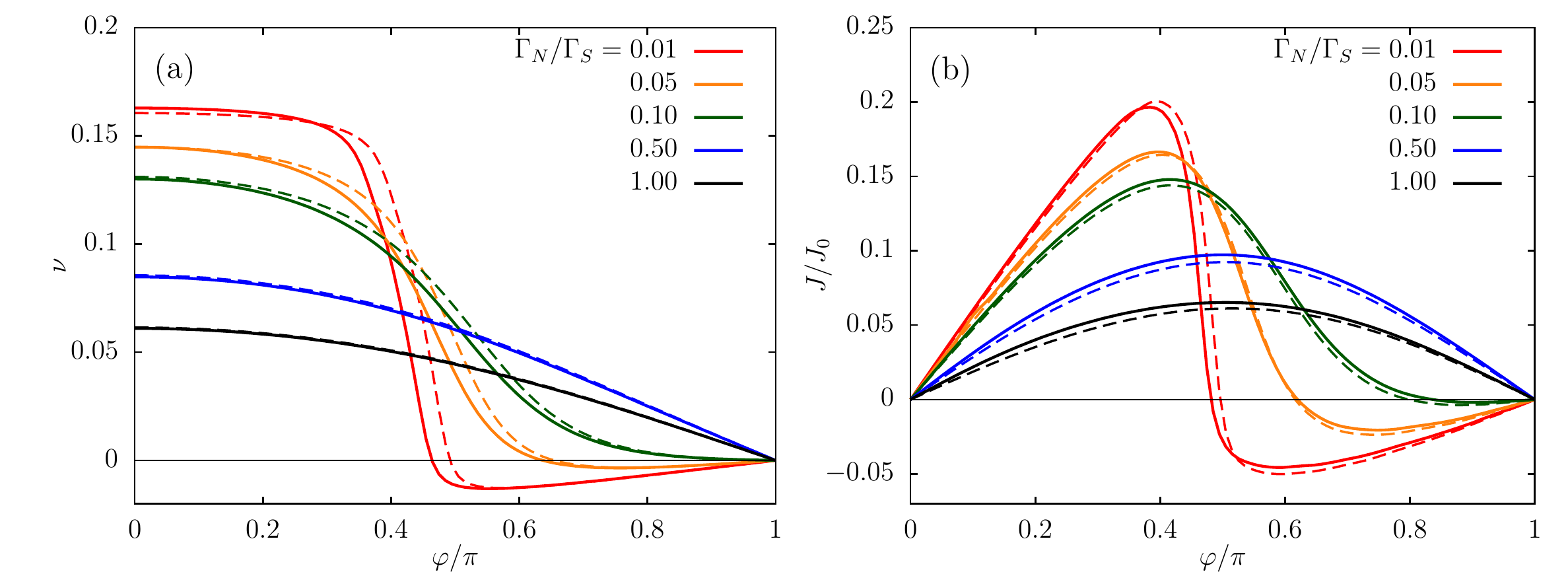}
	\caption{
		$(a)$ Phase evolution of the on-dot induced pairing $\nu\equiv\langle d_{\downarrow}d_{\uparrow} \rangle $ of the finite-gap three-terminal setup at $T=0$ obtained using NRG Ljubljana in the $w$ basis with subsequent use of Eq.~(\ref{PairingMap}) for the half-bandwidth $B=2000\Delta$ (solid lines) and $B=100\Delta$ (dashed lines). Parameters of the model are  $U=3\Delta$ and $\Gamma_S=\Delta$ while $\Gamma_N$ varies. 
		$(b)$ Phase evolution of the Josephson current $J$ with $J_0=2e\Delta/\hbar$ obtained using NRG Ljubljana in the $w$ basis with subsequent use of Eq.~(\ref{JosephsonMap}) for the same parameters as in panel $(a)$.  \label{fig_8}  }
\end{figure*}

To obtain the Josephson current $J$ we note that the corresponding operator involves $c$ electrons of the leads and we therefore use its expression in terms of the anomalous Green function $G_a^D(\omega^+)$ in the $d$ basis \cite{Zonda-2015}
\begin{equation}
J
=
2 \tan\frac{\varphi}{2}
\int_{-\infty}^{+\infty}\!\!
\frac{d\omega}{\pi}
f(\omega)
\mathrm{Im}
\left[
G_a^D(\omega^+) \Sigma^D_a(\omega^+)
\right],
\label{JosephsonMap}
\end{equation}
where  $f(\omega)$ is the Fermi-Dirac distribution,  while the anomalous self-energy %\eqref{eq:self-energy} 
$\Sigma^D_{a}(\omega^+)$ reads in the limit of the infinite bandwidth
\begin{eqnarray}
\Sigma^D_a(\omega^+)
&=&
\frac{\Gamma_S \Delta \cos(\varphi/2)}{\sqrt{\Delta^2-\omega^2}}
\Theta(\Delta^2-\omega^2)
\nonumber
\\
&+&
\frac{i\Gamma_S \Delta \cos(\varphi/2) \, \mathrm{sgn}(\omega)}{\sqrt{\omega^2-\Delta^2}}
\Theta(\omega^2-\Delta^2).
\end{eqnarray}
The integral (\ref{JosephsonMap}) involving anomalous components of the Green function requires high frequency resolution and reliable broadening procedure when applying NRG.

The results of phase-dependent Josephson current are shown in Fig.~\ref{fig_8} for $U=3\Delta$ at different ratios of $\Gamma_N/\Gamma_S$. The sign reversal of the Josephson current occurs only for small hybridization strengths $\Gamma_N$ which clearly shows that Kondo correlations are important in the system and may overcome the superconducting correlations. Once again, the $\varphi^*_{j}$ at which sign reversal occurs does not match $\varphi^*$ from crossing of the ABS states nor $\varphi^*_{\mathrm{pair}}$, but all three values tend to be the same for $\Gamma_N \rightarrow 0$.

\section{Conclusions \label{sec_5}} 

We have investigated a general finite-gap model of QD with an arbitrary Coulomb repulsion attached to the hybrid reservoir composed of one normal lead and two BCS leads with an arbitrary phase difference. To obtain reliable and method-unbiased results on phase-dependent spectral functions, the standard NRG was employed. However, the full problem with three types of leads requires in the standard NRG approach the implementation of three-channel calculations which poses several non-trivial challenges \cite{Mitchell-2014}. To circumvent the numerical limitations we have thus introduced a unitary transformation $\mathbb{T}$ of the local dot electrons $d$, which despite the general belief \cite{Hecht-2008} allows to reformulate the present finite-gap three-terminal model as well as any general model with phase-biased superconductors involved, including SCIAM, as a one channel problem. %Moreover, the approach holds also in the case of the decoupled normal electrode where the three-terminal setup goes over into the ordinary SCIAM. Consequently, at the half-filling $\mathbb{T}$ transforms SCIAM onto the SIAM with scalar but gapped TDOS. The vanishing TDOS around the Fermi energy consequently hampers the formation of the Kondo resonance. However, the quantitative understanding requires NRG techniques with either finite-sized Wilson chains, when the standard logarithmic discretization is used, or to modify the discretization to avoid the energy range of the hard gap. 

Since the present three-terminal reservoir has a non-zero TDOS around the Fermi energy a standard logarithmic discretization in the transformed basis of the fields $w_{\sigma}$ with $\sigma \in \{ \uparrow, \downarrow\}$ can be employed. Thus, the open-source NRG Ljubljana code could be employed unaltered. The obtained phase-dependent spectral functions showed behavior resembling that of the SCIAM, see  also Figs.~\ref{fig_5} and \ref{fig_8}. Thus, two regimes, referred here as the $0$-like and the $\pi$-like phase, have been identified in analogy. They do not however constitute separate phases since the non-zero TDOS around the Fermi energy leads to the formation of a singlet many-body ground state for any $\varphi$.

The width of the resulting crossover region, see Figs.~\ref{fig_5} and \ref{fig_8}, is roughly proportional to the hybridization $\Gamma_N$. Here, the off-center in-gap peaks of the corresponding spectral function do cross at $\varphi^*$, the on-dot induced pairing changes sign at $\varphi^*_{\mathrm{pair}}$ and the Josephson current at $\varphi^*_{j}$. Generally the values of $\varphi^*$,  $\varphi^*_{\mathrm{pair}}$ and  $\varphi^*_{j}$ do not equal, but do so in the limit of $\Gamma_N \rightarrow 0$, where the corresponding SCIAM limit is obtained. However, in the limit of large $\Gamma_N$ pairing and Josephson current are always positive and neither $\varphi^*_{\mathrm{pair}}$ nor $\varphi^*_{j}$ are defined although $\varphi^*$ still exists as the $0$-like and the $\pi$-like distinct spectra are present.

Thus, for $\varphi>\varphi^*$, the presence of two pairs of the in-gap peaks which merge together at $\varphi=\pi$ and the presence of a central Kondo-like resonance defines the spectral property of the $\pi$-like phase. The two pairs of the in-gap peaks show an analogous phase-dependent behavior as the ABS states of the SCIAM as shown in Fig.~\ref{fig_5} and they can consequently be understood as the broadened analogs of the ABS states of the SCIAM. However, the non-zero TDOS around the Fermi energy, as provided by the normal lead, allows screening of the spin of the QD even at $T=0$ and  leads to the formation of the Kondo peak in the spectral function of the $\pi$-like phase. Thus, unlike in the SCIAM, in the finite-gap three-terminal setup the broadened ABS states do co-exist with the Kondo resonance in the $\pi$-like phase as shown in panels $(a)$ and $(c)$ of Fig.~\ref{fig_5}. 

Nevertheless, at $\varphi^*$ the central peak at the Fermi energy splits and can no longer be attributed to Kondo-like correlations since charge excitations dominate the effective underlying model which is strongly out of the half-filling. Thus, for $\varphi<\varphi^*$ the $0$-like region is entered with the spectral weight at the Fermi energy moving towards zero with further decreasing $\varphi$. The resulting split peak can then be interpreted in terms of two broadened ABS states of phase-dependent behavior resembling the SCIAM. However, such a $0$-like phase has an additional pair of low-intensity peaks at higher frequencies, which (unlike in SCIAM) can be excited in the one-particle manner due to admixtures of the doublet state induced by the coupling to the normal lead.

Such a complex behavior is qualitatively explained via the transformation $\mathbb{T}$ which is thus not merely a technical tool for the NRG implementation. The TDOS in the $w$ basis is highly particle-hole asymmetric at $\varphi=0$. Then, with increasing $\varphi$, its asymmetry is continuously diminished until at $\varphi=\pi$ it completely vanishes. As shown in Fig.~\ref{fig_4}, the particle-hole asymmetry of the TDOS effectively acts as the particle-hole asymmetry in an ordinary SIAM with the concomitant increase of the Kondo temperature upon increasing the asymmetry followed by entering the mixed valence regime and eventually complete destruction of the Kondo resonance. 

The Kondo temperature $T_K$ can be quantitatively assessed  via the phase-dependent HWHM of the central Kondo-like peak. The analysis in Sec.~\ref{subsec_transp} (see also Fig.~\ref{fig_6}) showed that almost up to $\varphi^*$, the $\log T_K$ follows very well the $\cos^2(\varphi/2)$ trend hypothesized already in Ref.~\cite{Domanski-2017} for $\Gamma_N \lesssim \Delta/10$. Thus, even though the hypothesis in Ref.~\cite{Domanski-2017} is based on the infinite-gap limit of the present model and the second-order perturbation theory, we have shown that it is robust and holds for sufficiently weak $\Gamma_N$. Significant deviations from the $\cos^2(\varphi/2)$ law start appearing roughly around $\Gamma_N \sim \Delta$.  

Moreover, using the transformation $\mathbb{T}$ we have also obtained the phase-dependent on-dot induced pairing $\nu$ and the phase-dependent Josephson current $J$ in various parametric ranges. Results in the limit of the infinitely wide band are presented in Fig.~\ref{fig_8}, where also effects of finite width of the band are shown (with a detailed derivation given in the Appendix~\ref{app_finite}). Incorporating these corrections allowed for comparison with another numerically exact method, the continuous-time hybridization expansion (CT-HYB) QMC with a good agreement in the regions outside of the crossover while large temperature dependence smears the region itself, see Appendix \ref{app_qmc}. The resulting pairing and supercurrent reversals do not occur exactly at $\varphi^*$ defined by the spectral functions and appear only at sufficiently low ratios $\Gamma_N/\Gamma_S$. Once a given threshold is exceeded and the Kondo screening dominates the system, superconducting correlations are essentially suppressed and only modify the phase-dependent transport. This behavior is enhanced by increasing the interaction strength, reducing the gap size, or increasing the hybridization of the normal lead. The observation of the $0$-$\pi$-like crossover is thus possible only in a fairly small portion of the parameter space corresponding to the weak coupling of the normal lead to the QD.

The mapping $\mathbb{T}$ not only significantly reduces the numerical complexity of the hybrid normal-superconductor reservoirs, but it also allows for conceptual understanding of the competing Kondo and Josephson effects via the particle-hole asymmetric SIAM. In this regard, it is worth mentioning that the transformation $\mathbb{T}$ applies in the same form also to the SCIAM, i.e., an interacting QD coupled to two superconducting leads with hard gap in the spectrum, leading to its mapping onto the problem of normal Anderson impurity coupled to an insulator-like electronic reservoir with a hard spectral gap around the Fermi level. The original Nambu formulation becomes then a scalar one which may allow for new insights and is thus worth further pursuits. Moreover, as shown in Ref.~\cite[Fig.~3b]{Kadlecova-2019}, in the most interesting Kondo regime of the SCIAM model the results are in fact independent of the value of the particle-hole asymmetry for quite a wide range of its value. Therefore, even though the mapping $\mathbb{T}$ is restricted to the particle-hole symmetric model, the obtained results should be applicable also rather far away from this regime which makes the transformed scalar version of the SCIAM model practically relevant.

\begin{acknowledgments}
We acknowledge discussions with Rok \v{Z}itko, Martin \v{Z}onda, and V\'{a}clav Jani\v{s}.
This work was supported by Grant  No. 19-13525S of the Czech Science Foundation (PZ, TN), 
by grant INTER-COST LTC19045 (VP), 
by the COST Action NANOCOHYBRI (CA16218) (TN),
the National Science Centre (NCN, Poland) via Grant No.\ UMO-2017/27/B/ST3/01911 (TN) and
by The Ministry of Education, Youth and Sports from the Large Infrastructures for Research, 
Experimental Development and Innovations project 
``IT4Innovations National Supercomputing Center – LM2015070'' and project 
"e-Infrastruktura CZ" (e-INFRA LM2018140).
%Access to computing  and  storage facilities owned by parties and projects 
%contributing to the National Grid Infrastructure MetaCentrum provided under 
%the programme ``Projects of Large Research, Development, and Innovations Infrastructures'' 
%(CESNET LM2015042) is also appreciated.

\end{acknowledgments}

\appendix

\section{Interaction term in the $\bf w$ basis \label{app_hubbard}}

\begin{figure*}[ht]
	\includegraphics[width=2.00\columnwidth]{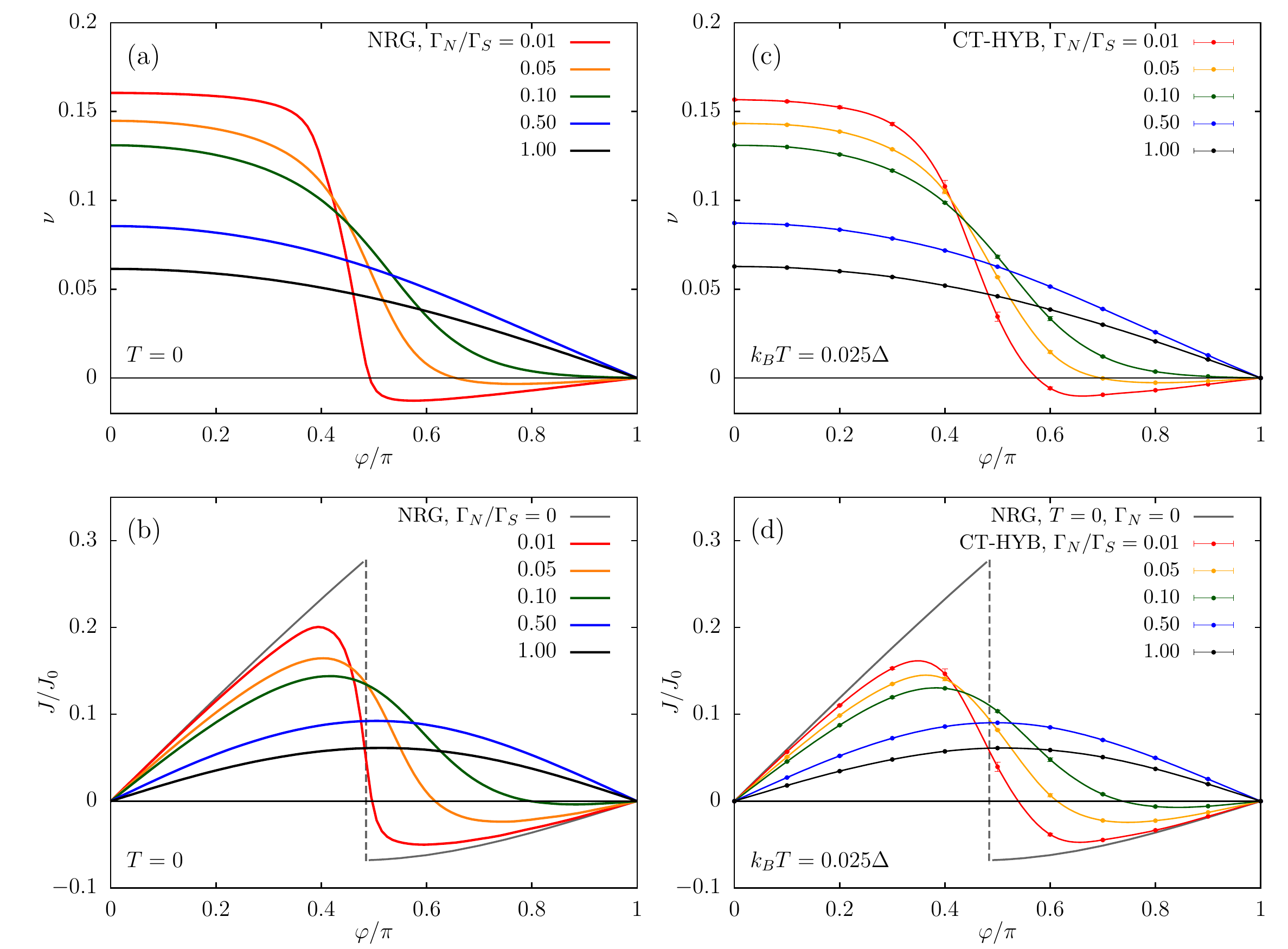}
	\caption{
		Left column: Phase-dependent on-dot pairing $\nu\equiv\langle d_{\downarrow}d_{\uparrow} \rangle$ (panel $a$) and the Josephson current $J$ with $J_0=2e\Delta/\hbar$ (panel b)
		calculated using NRG at zero temperature in the  $w$ fields with subsequent use of Eqs.~\eqref{PairingMap} and \eqref{JosephsonMap} for the parameters 
		$B=100\Delta$, $U=3\Delta$, $\Gamma_S=\Delta$ and various values of $\Gamma_N$.
		Right column: Equivalent results calculated using CT-HYB QMC in the basis of the $d$ fields for small finite temperature $k_BT=\Delta/40$ (symbols with error bars):
		on-dot pairing $\nu$ (panel $c$) and Josephson current $J$ (panel d). Lines are splines of QMC data and serve only as a guide to the eye.
		\label{fig_9} }
\end{figure*}

The expression (\ref{Hub_w}) is easily obtained by first applying the transformation $\mathbb{T}^-_1$ to the following quantities
\begin{eqnarray}
d_{\uparrow}^{\dagger}
d_{\uparrow}^{\vphantom{\dagger}}
&=&
\frac{
1
+ 
w_{\uparrow}^{\dagger}
w_{\uparrow}^{\vphantom{\dagger}}
- 
w_{\downarrow}^{\dagger}
w_{\downarrow}^{\vphantom{\dagger}}
}{2}
-
\frac{
w_{\uparrow}^{\dagger}
w_{\downarrow}^{\dagger}
+ 
w_{\downarrow}^{\vphantom{\dagger}} 
w_{\uparrow}^{\vphantom{\dagger}}
}{2},
\\
d_{\downarrow}^{\dagger}
d_{\downarrow}^{\vphantom{\dagger}}
&=&
\frac{
1
-
w_{\uparrow}^{\dagger}
w_{\uparrow}^{\vphantom{\dagger}}
+ 
w_{\downarrow}^{\dagger}
w_{\downarrow}^{\vphantom{\dagger}}
}{2}
-
\frac{
w_{\uparrow}^{\dagger}
w_{\downarrow}^{\dagger}
+ 
w_{\downarrow}^{\vphantom{\dagger}}
w_{\uparrow}^{\vphantom{\dagger}}
}{2}.
\end{eqnarray}
Since
$\mu = w_{\uparrow}^{\dagger} w_{\uparrow}^{\vphantom{\dagger}} - w_{\downarrow}^{\dagger} w_{\downarrow}^{\vphantom{\dagger}} $ 
and 
$\xi = w_{\uparrow}^{\dagger} w_{\downarrow}^{\dagger}
+ w_{\downarrow}^{\vphantom{\dagger}} w_{\uparrow}^{\vphantom{\dagger}}
$ satisfy
\begin{eqnarray}
1 - \mu^2 
&=&
\xi^2 
=
- 
w_{\uparrow}^{\dagger}
w_{\uparrow}^{\vphantom{\dagger}}
+ 
2 n_{w\uparrow} n_{w\downarrow}
+ 
w_{\downarrow}^{\vphantom{\dagger}}
w_{\downarrow}^{\dagger},
\\
\mu \xi 
&=&
\xi \mu
=
0,
\end{eqnarray}
with
$n_{w\uparrow} = w_{\uparrow}^{\dagger}
w_{\uparrow}^{\vphantom{\dagger}}$
and 
$n_{w\downarrow} = w_{\downarrow}^{\dagger}
w_{\downarrow}^{\vphantom{\dagger}}$,
we obtain
\begin{eqnarray}
d_{\uparrow}^{\dagger}
d_{\uparrow}^{\vphantom{\dagger}}
d_{\downarrow}^{\dagger}
d_{\downarrow}^{\vphantom{\dagger}}
&=&
\frac{
1 -\mu^2 - \mu \xi + \xi \mu -2 \xi + \xi^2
}{4}
\nonumber \\
&=&
n_{w\uparrow} n_{w\downarrow} 
-
\frac{1}{2}
W^{\dagger}
\left(
\sigma_x + \sigma_z
\right)
W^{\vphantom{\dagger}}.
\label{A5}
\end{eqnarray}
Applying then $\mathbb{T}^-_1$ to the interaction term $H_U$ gives
\begin{eqnarray}
H_U
&=&
U
d^{\dagger}_{\uparrow}
d^{\vphantom{\dagger}}_{ \uparrow}
d^{\dagger}_{\downarrow}
d^{\vphantom{\dagger}}_{ \downarrow}
-\frac{U}{2}
D^{\dagger} 
\left(
\sigma_x+\sigma_z
\right)
D^{\vphantom{\dagger}}
\nonumber \\
&=&
U
w^{\dagger}_{\uparrow}
w^{\vphantom{\dagger}}_{ \uparrow}
w^{\dagger}_{\downarrow}
w^{\vphantom{\dagger}}_{ \downarrow},
\end{eqnarray}
which in the $w$ basis obtains the form of the ordinary Hubbard term. Notice that the additional quadratic term in $H_U$ in the $d$ basis cancels exactly the quadratic term of Eq.~(\ref{A5}) when correspondingly transformed.

\section{Corrections due to the finite bandwidth \label{app_finite} }

The self-energy contribution $\mathbb{\Sigma}^D$ can be written a sum
\begin{equation}
\mathbb{\Sigma}^D(z) 
=
\mathbb{\Sigma}^D_N(z) 
+
\mathbb{\Sigma}^D_S(z)\ ,
\end{equation}
where $\mathbb{\Sigma}^D_N$ is the contribution of just the normal lead and $\mathbb{\Sigma}^D_S$ is the same due to the superconducting leads:
\begin{eqnarray}
\mathbb{\Sigma}^D_N(z)
&=&
\sum_{\mathbf{k}}
\mathbb{V}_{N\mathbf{k}}
\left(
z \cdot \mathbb{1} - \mathbb{E}_{N\mathbf{k}}
\right)^{-1}
\mathbb{V}_{N\mathbf{k}}
\nonumber
\\
\mathbb{\Sigma}^D_S(z)
&=&
\!\!\!\!\!\!\!
\sum_{\alpha \in \{L,R\},\mathbf{k}}
\!\!\!\!\!\!\!\!\mathbb{V}_{\alpha\mathbf{k}}
\left(
z \cdot \mathbb{1} - \mathbb{E}_{\alpha\mathbf{k}}
\right)^{-1}
\mathbb{V}_{\alpha\mathbf{k}}
\end{eqnarray}
with $z$ being an arbitrary complex number. Later, only the functional form only infinitesimally close to the real axis needs to be resolved. For that we set  $z=\omega^+\equiv \omega+i\eta $ with $\omega$ being a real frequency and $\eta$ being an infinitesimally small positive number, thus taking the cut slightly above the real axis. The self-energy contribution $\mathbb{\Sigma}^D_N(\omega^+)$ for the constant TDOS within the band \eqref{TDOS} simply reads 
\begin{equation}
\mathbb{\Sigma}^D_N(\omega^+)
=
-i\Gamma_N \mathbb{1}
\text{ , for } |\omega|<B.
\end{equation}
The superconducting part is non-trivial and shall be treated here in more detail. It is defined as
\begin{equation}
\mathbb{\Sigma}^D_S(\omega^+)
=
\sum_{\alpha \in \{L,R\}\mathbf{k}}
\!\!\!\!\!\mathbb{V}_{\alpha\mathbf{k}}
\left(
 \omega^+ \mathbb{1} - \mathbb{E}_{\alpha\mathbf{k}}
\right)^{-1}
\mathbb{V}_{\alpha\mathbf{k}}
\label{B4}
\end{equation}
with notation following Sec.~\ref{subsec_nambu}. The inverse matrix appearing in Eq.~(\ref{B4}) is evaluated using the identity $(u \mathbb{1}+\vec{v}\cdot\vec{\sigma})^{-1}=(u \mathbb{1}-\vec{v}\cdot\vec{\sigma})/(u^2-\vec{v}\cdot\vec{v})$ as
\begin{equation}
\left(
\omega^+ \mathbb{1} - \mathbb{E}_{\alpha\mathbf{k}}
\right)^{-1}
= \frac{
	\omega \mathbb{1} 
	- \Delta C_{\alpha} \sigma_x
	+ \Delta S_{\alpha} \sigma_y
	+\varepsilon_{\mathbf{k}\alpha} \sigma_z
	}{(\omega+i\eta)^2 - \Delta^2 - \varepsilon_{\mathbf{k}\alpha}^2}.
\end{equation}
Furthermore, since $\sigma_z(u \mathbb{1}+v_x\sigma_x+v_y\sigma_y+v_z\sigma_z)\sigma_z=u \mathbb{1}-v_x\sigma_x-v_y\sigma_y+v_z\sigma_z$, we get
\begin{eqnarray}
\mathbb{V}_{\alpha\mathbf{k}}
\left(
 \omega^+ \mathbb{1} - \mathbb{E}_{\alpha\mathbf{k}}
\right)^{-1}
\mathbb{V}_{\alpha\mathbf{k}}=
\hspace{3cm}
\nonumber
\\
V_{\alpha\mathbf{k}}^2
\frac{
	\omega \mathbb{1} 
	+ \Delta C_{\alpha} \sigma_x
	- \Delta S_{\alpha} \sigma_y
	+\varepsilon_{\mathbf{k}\alpha} \sigma_z
	}
	{\omega^2 - \Delta^2 - \varepsilon_{\mathbf{k}\alpha}^2 + i\eta\, \mathrm{sgn}(\omega)
	},
\label{B6}
\end{eqnarray}
which, eventually, under the assumption \eqref{TDOS} of constant TDOS within the band leads to
\begin{equation}
\mathbb{\Sigma}^D_S(\omega^+)
=
\!\! \sum_{\alpha \in \{L,R\}} \!\!
\frac{\Gamma_{\alpha}}{\pi}
\int_{-B}^{B}
\frac{
	\omega \mathbb{1} 
	+ \Delta C_{\alpha} \sigma_x
	- \Delta S_{\alpha} \sigma_y
}{\omega^2 - \Delta^2 - \varepsilon^2 + i\eta\, \mathrm{sgn}(\omega)}
d\varepsilon,
\end{equation}
where the term proportional to $\sigma_z$ vanished due to the integrand being an odd function of $\varepsilon$. Using the symmetric phase drop gauge choice (cf.~the discussion in Sec.~\ref{subsec_micro}) $\varphi_L=-\varphi_R=\varphi/2$, we sum over $\alpha \in \{ L, R\}$ yielding
\begin{equation}
\mathbb{\Sigma}^D_S(\omega^+)
=
\Gamma_S
\left[	
\omega \mathbb{1} 
+ \Delta \cos\left(\frac{\varphi}{2}\right) \sigma_x
\right]
F(\omega^+)\ ,
\end{equation}
with
\begin{equation}
\begin{split}
F(\omega^+)
&\equiv
\frac{1}{\pi}
\int_{-B}^{B}
\frac{d\varepsilon}{\omega^2 - \Delta^2 - \varepsilon^2 + i\eta\, \mathrm{sgn}(\omega)}\\
&=\frac{1}{\pi\sqrt{(\omega+ i\eta)^2 - \Delta^2 }}\ln\frac{\sqrt{(\omega+ i\eta)^2 - \Delta^2}+B}{\sqrt{(\omega+ i\eta)^2 - \Delta^2}-B}.
\end{split}
\end{equation}

Taking the $\eta \rightarrow 0$ limit, we arrive at 
\begin{equation}
F(\omega^+)
=
\begin{cases}
-\frac{2}{\pi\sqrt{\Delta^2-\omega^2}} \arctan
\left(\frac{B}{\sqrt{\Delta^2-\omega^2}}\right),
&
\text{for }|\omega|<\Delta
\\
-\frac{i\, \mathrm{sgn}(\omega)}{\sqrt{\omega^2-\Delta^2}}
+
\frac{\ln \left( \frac{B+\sqrt{\omega^2-\Delta^2}}{B-\sqrt{\omega^2-\Delta^2}}\right)}{\pi\sqrt{\omega^2-\Delta^2}},
&
\text{for }\Delta<|\omega|<B.
\end{cases}
\end{equation}
The resulting $\mathbb{\Sigma}^D_S(\omega^+)$ has thus a non-zero imaginary part only outside of the gap region while all effects of the finite-sized band appear in its real part which is non-zero in the whole band. However, once the limit $B\rightarrow\infty$ is taken the real 
part out of the gap vanishes too. 

Altogether, the self-energy contribution $\mathbb{\Sigma}^D(\omega^+)$ takes the form of Eq.~(\ref{eq:self-energy}), where
\begin{eqnarray}
\Sigma^D_n(\omega^+)
&=&
-i\Gamma_N
+ \,
\Gamma_S
\omega
F(\omega^+)
\\
\Sigma^D_a(\omega^+)
&=&
\Gamma_S\Delta\cos\left(\frac{\varphi}{2}\right)
F(\omega^+).
\end{eqnarray}

\section{Comparison of the NRG results with QMC \label{app_qmc}}

In order to assess the ability of the presented NRG scheme to provide reliable results on the integral quantities 
like the on-dot induced pairing and the Josephson current, we compare the results with a numerically 
exact continuous-time, hybridization-expansion (CT-HYB) QMC, as this method was already successfully 
used to study both the two-terminal~\cite{Kadlecova-2019} and 
three-terminal~\cite{Domanski-2017} setups and agrees with standard NRG results well within the QMC error bars.

The CT-HYB calculation is performed in the original $d$ basis by employing the off-diagonal elements of the
hybridization function using the TRIQS/CTHYB solver~\cite{Seth-2016}. 
The total Hamiltonian of the system does not conserve particle number, therefore the superconducting pairing is introduced 
to the method using a canonical particle-hole transformation in the spin-down sector, mapping the system to 
an impurity Anderson model with attractive interaction~\cite{Luitz-2010,Pokorny-2018}. 
As CT-HYB is an inherently finite-temperature method, all calculations were performed at $k_BT=\Delta/40$.
All results are calculated for half-bandwidth $B=100\Delta$ and a cutoff in Matsubara frequencies 
$\omega_{n}^{max}\approx314\Delta$.

The comparison of the NRG results with CT-HYB method is plotted in Fig~\ref{fig_9}. In panel $a$ (top left), NRG results for the induced pairing $\nu$ as a function of phase difference $\varphi$ for $B=100\Delta$, $U=3\Delta$, $\Gamma_S=\Delta$ and various values of $\Gamma_N$ at $T=0$ are plotted. The importance of the finite-bandwidth corrections were already discussed in Fig.~\ref{fig_8}. The equivalent results of CT-HYB for small finite temperature are plotted in panel $c$ (top right). The curves match within QMC error bars for small and large values of $\varphi$. In the crossover region, the results slightly differ as the finite temperature is a source of additional smearing, having a similar effect as $\Gamma_N$ \cite{Domanski-2017}. In panel $b$ (bottom left) we plotted the NRG results for the Josephson current for the same set of parameters as in panel $a$. We added a $\Gamma_N=0$ result from Ref.~\cite{Domanski-2017} to mark the position of the QPT in a case of detached normal electrode. The relevant CT-HYB result is again plotted in panel $d$ (bottom right). Comparison again shows good agreement up to the finite-temperature effects.

%\bibliography{josephson}{}

\begin{thebibliography}{0}%
\makeatletter
\providecommand \@ifxundefined [1]{%
 \@ifx{#1\undefined}
}%
\providecommand \@ifnum [1]{%
 \ifnum #1\expandafter \@firstoftwo
 \else \expandafter \@secondoftwo
 \fi
}%
\providecommand \@ifx [1]{%
 \ifx #1\expandafter \@firstoftwo
 \else \expandafter \@secondoftwo
 \fi
}%
\providecommand \natexlab [1]{#1}%
\providecommand \enquote  [1]{``#1''}%
\providecommand \bibnamefont  [1]{#1}%
\providecommand \bibfnamefont [1]{#1}%
\providecommand \citenamefont [1]{#1}%
\providecommand \href@noop [0]{\@secondoftwo}%
\providecommand \href [0]{\begingroup \@sanitize@url \@href}%
\providecommand \@href[1]{\@@startlink{#1}\@@href}%
\providecommand \@@href[1]{\endgroup#1\@@endlink}%
\providecommand \@sanitize@url [0]{\catcode `\\12\catcode `\$12\catcode
  `\&12\catcode `\#12\catcode `\^12\catcode `\_12\catcode `\%12\relax}%
\providecommand \@@startlink[1]{}%
\providecommand \@@endlink[0]{}%
\providecommand \url  [0]{\begingroup\@sanitize@url \@url }%
\providecommand \@url [1]{\endgroup\@href {#1}{\urlprefix }}%
\providecommand \urlprefix  [0]{URL }%
\providecommand \Eprint [0]{\href }%
\providecommand \doibase [0]{http://dx.doi.org/}%
\providecommand \selectlanguage [0]{\@gobble}%
\providecommand \bibinfo  [0]{\@secondoftwo}%
\providecommand \bibfield  [0]{\@secondoftwo}%
\providecommand \translation [1]{[#1]}%
\providecommand \BibitemOpen [0]{}%
\providecommand \bibitemStop [0]{}%
\providecommand \bibitemNoStop [0]{.\EOS\space}%
\providecommand \EOS [0]{\spacefactor3000\relax}%
\providecommand \BibitemShut  [1]{\csname bibitem#1\endcsname}%
\let\auto@bib@innerbib\@empty
%</preamble>
\end{thebibliography}%


\begin{thebibliography}{59}%
	\makeatletter
	\providecommand \@ifxundefined [1]{%
		\@ifx{#1\undefined}
	}%
	\providecommand \@ifnum [1]{%
		\ifnum #1\expandafter \@firstoftwo
		\else \expandafter \@secondoftwo
		\fi
	}%
	\providecommand \@ifx [1]{%
		\ifx #1\expandafter \@firstoftwo
		\else \expandafter \@secondoftwo
		\fi
	}%
	\providecommand \natexlab [1]{#1}%
	\providecommand \enquote  [1]{``#1''}%
	\providecommand \bibnamefont  [1]{#1}%
	\providecommand \bibfnamefont [1]{#1}%
	\providecommand \citenamefont [1]{#1}%
	\providecommand \href@noop [0]{\@secondoftwo}%
	\providecommand \href [0]{\begingroup \@sanitize@url \@href}%
	\providecommand \@href[1]{\@@startlink{#1}\@@href}%
	\providecommand \@@href[1]{\endgroup#1\@@endlink}%
	\providecommand \@sanitize@url [0]{\catcode `\\12\catcode `\$12\catcode
		`\&12\catcode `\#12\catcode `\^12\catcode `\_12\catcode `\%12\relax}%
	\providecommand \@@startlink[1]{}%
	\providecommand \@@endlink[0]{}%
	\providecommand \url  [0]{\begingroup\@sanitize@url \@url }%
	\providecommand \@url [1]{\endgroup\@href {#1}{\urlprefix }}%
	\providecommand \urlprefix  [0]{URL }%
	\providecommand \Eprint [0]{\href }%
	\providecommand \doibase [0]{http://dx.doi.org/}%
	\providecommand \selectlanguage [0]{\@gobble}%
	\providecommand \bibinfo  [0]{\@secondoftwo}%
	\providecommand \bibfield  [0]{\@secondoftwo}%
	\providecommand \translation [1]{[#1]}%
	\providecommand \BibitemOpen [0]{}%
	\providecommand \bibitemStop [0]{}%
	\providecommand \bibitemNoStop [0]{.\EOS\space}%
	\providecommand \EOS [0]{\spacefactor3000\relax}%
	\providecommand \BibitemShut  [1]{\csname bibitem#1\endcsname}%
	\let\auto@bib@innerbib\@empty
	%</preamble>
	\bibitem [{\citenamefont {Tans}\ \emph {et~al.}(1997)\citenamefont {Tans},
		\citenamefont {Devoret}, \citenamefont {Dai}, \citenamefont {Thess},
		\citenamefont {Smalley}, \citenamefont {Geerligs},\ and\ \citenamefont
		{Dekker}}]{Tans-1997}%
	\BibitemOpen
	\bibfield  {author} {\bibinfo {author} {\bibfnamefont {S.~J.}\ \bibnamefont
			{Tans}}, \bibinfo {author} {\bibfnamefont {M.~H.}\ \bibnamefont {Devoret}},
		\bibinfo {author} {\bibfnamefont {H.}~\bibnamefont {Dai}}, \bibinfo {author}
		{\bibfnamefont {A.}~\bibnamefont {Thess}}, \bibinfo {author} {\bibfnamefont
			{R.~E.}\ \bibnamefont {Smalley}}, \bibinfo {author} {\bibfnamefont {L.~J.}\
			\bibnamefont {Geerligs}}, \ and\ \bibinfo {author} {\bibfnamefont
			{C.}~\bibnamefont {Dekker}},\ }\href {http://dx.doi.org/10.1038/386474a0}
	{\bibfield  {journal} {\bibinfo  {journal} {Nature}\ }\textbf {\bibinfo
			{volume} {386}},\ \bibinfo {pages} {474} (\bibinfo {year}
		{1997})}\BibitemShut {NoStop}%
	\bibitem [{\citenamefont {Kasumov}\ \emph {et~al.}(1999)\citenamefont
		{Kasumov}, \citenamefont {Deblock}, \citenamefont {Kociak}, \citenamefont
		{Reulet}, \citenamefont {Bouchiat}, \citenamefont {Khodos}, \citenamefont
		{Gorbatov}, \citenamefont {Volkov}, \citenamefont {Journet},\ and\
		\citenamefont {Burghard}}]{Kasumov-1999}%
	\BibitemOpen
	\bibfield  {author} {\bibinfo {author} {\bibfnamefont {A.~Y.}\ \bibnamefont
			{Kasumov}}, \bibinfo {author} {\bibfnamefont {R.}~\bibnamefont {Deblock}},
		\bibinfo {author} {\bibfnamefont {M.}~\bibnamefont {Kociak}}, \bibinfo
		{author} {\bibfnamefont {B.}~\bibnamefont {Reulet}}, \bibinfo {author}
		{\bibfnamefont {H.}~\bibnamefont {Bouchiat}}, \bibinfo {author}
		{\bibfnamefont {I.~I.}\ \bibnamefont {Khodos}}, \bibinfo {author}
		{\bibfnamefont {Y.~B.}\ \bibnamefont {Gorbatov}}, \bibinfo {author}
		{\bibfnamefont {V.~T.}\ \bibnamefont {Volkov}}, \bibinfo {author}
		{\bibfnamefont {C.}~\bibnamefont {Journet}}, \ and\ \bibinfo {author}
		{\bibfnamefont {M.}~\bibnamefont {Burghard}},\ }\href
	{http://www.sciencemag.org/content/284/5419/1508} {\bibfield  {journal}
		{\bibinfo  {journal} {Science}\ }\textbf {\bibinfo {volume} {284}},\ \bibinfo
		{pages} {1508} (\bibinfo {year} {1999})}\BibitemShut {NoStop}%
	\bibitem [{\citenamefont {Jarillo-Herrero}\ \emph {et~al.}(2006)\citenamefont
		{Jarillo-Herrero}, \citenamefont {van Dam},\ and\ \citenamefont
		{Kouwenhoven}}]{Jarillo-2006}%
	\BibitemOpen
	\bibfield  {author} {\bibinfo {author} {\bibfnamefont {P.}~\bibnamefont
			{Jarillo-Herrero}}, \bibinfo {author} {\bibfnamefont {J.~A.}\ \bibnamefont
			{van Dam}}, \ and\ \bibinfo {author} {\bibfnamefont {L.~P.}\ \bibnamefont
			{Kouwenhoven}},\ }\href {http://dx.doi.org/10.1038/nature04550} {\bibfield
		{journal} {\bibinfo  {journal} {Nature}\ }\textbf {\bibinfo {volume} {439}},\
		\bibinfo {pages} {953} (\bibinfo {year} {2006})}\BibitemShut {NoStop}%
	\bibitem [{\citenamefont {J{\o}rgensen}\ \emph {et~al.}(2006)\citenamefont
		{J{\o}rgensen}, \citenamefont {Grove-Rasmussen}, \citenamefont {Novotn{\'y}},
		\citenamefont {Flensberg},\ and\ \citenamefont {Lindelof}}]{Jorgensen-2006}%
	\BibitemOpen
	\bibfield  {author} {\bibinfo {author} {\bibfnamefont {H.~I.}\ \bibnamefont
			{J{\o}rgensen}}, \bibinfo {author} {\bibfnamefont {K.}~\bibnamefont
			{Grove-Rasmussen}}, \bibinfo {author} {\bibfnamefont {T.}~\bibnamefont
			{Novotn{\'y}}}, \bibinfo {author} {\bibfnamefont {K.}~\bibnamefont
			{Flensberg}}, \ and\ \bibinfo {author} {\bibfnamefont {P.~E.}\ \bibnamefont
			{Lindelof}},\ }\href {http://link.aps.org/doi/10.1103/PhysRevLett.96.207003}
	{\bibfield  {journal} {\bibinfo  {journal} {Phys. Rev. Lett.}\ }\textbf
		{\bibinfo {volume} {96}},\ \bibinfo {pages} {207003} (\bibinfo {year}
		{2006})}\BibitemShut {NoStop}%
	\bibitem [{\citenamefont {Cleuziou}\ \emph {et~al.}(2006)\citenamefont
		{Cleuziou}, \citenamefont {Wernsdorfer}, \citenamefont {Bouchiat},
		\citenamefont {Ondarcuhu},\ and\ \citenamefont {Monthioux}}]{Cleuziou-2006}%
	\BibitemOpen
	\bibfield  {author} {\bibinfo {author} {\bibfnamefont {J.~P.}\ \bibnamefont
			{Cleuziou}}, \bibinfo {author} {\bibfnamefont {W.}~\bibnamefont
			{Wernsdorfer}}, \bibinfo {author} {\bibfnamefont {V.}~\bibnamefont
			{Bouchiat}}, \bibinfo {author} {\bibfnamefont {T.}~\bibnamefont {Ondarcuhu}},
		\ and\ \bibinfo {author} {\bibfnamefont {M.}~\bibnamefont {Monthioux}},\
	}\href {http://dx.doi.org/10.1038/nnano.2006.54} {\bibfield  {journal}
	{\bibinfo  {journal} {Nat. Nanotechnol.}\ }\textbf {\bibinfo {volume} {1}},\
	\bibinfo {pages} {53} (\bibinfo {year} {2006})}\BibitemShut {NoStop}%
\bibitem [{\citenamefont {Eichler}\ \emph {et~al.}(2009)\citenamefont
	{Eichler}, \citenamefont {Deblock}, \citenamefont {Weiss}, \citenamefont
	{Karrasch}, \citenamefont {Meden}, \citenamefont {Sch{\"o}nenberger},\ and\
	\citenamefont {Bouchiat}}]{Eichler-2009}%
\BibitemOpen
\bibfield  {author} {\bibinfo {author} {\bibfnamefont {A.}~\bibnamefont
		{Eichler}}, \bibinfo {author} {\bibfnamefont {R.}~\bibnamefont {Deblock}},
	\bibinfo {author} {\bibfnamefont {M.}~\bibnamefont {Weiss}}, \bibinfo
	{author} {\bibfnamefont {C.}~\bibnamefont {Karrasch}}, \bibinfo {author}
	{\bibfnamefont {V.}~\bibnamefont {Meden}}, \bibinfo {author} {\bibfnamefont
		{C.}~\bibnamefont {Sch{\"o}nenberger}}, \ and\ \bibinfo {author}
	{\bibfnamefont {H.}~\bibnamefont {Bouchiat}},\ }\href
{http://link.aps.org/doi/10.1103/PhysRevB.79.161407} {\bibfield  {journal}
	{\bibinfo  {journal} {Phys. Rev. B}\ }\textbf {\bibinfo {volume} {79}},\
	\bibinfo {pages} {161407} (\bibinfo {year} {2009})}\BibitemShut {NoStop}%
\bibitem [{\citenamefont {Pillet}\ \emph {et~al.}(2010)\citenamefont {Pillet},
	\citenamefont {Quay}, \citenamefont {Morfin}, \citenamefont {Bena},
	\citenamefont {Yeyati},\ and\ \citenamefont {Joyez}}]{Pillet-2010}%
\BibitemOpen
\bibfield  {author} {\bibinfo {author} {\bibfnamefont {J.-D.}\ \bibnamefont
		{Pillet}}, \bibinfo {author} {\bibfnamefont {C.~H.~L.}\ \bibnamefont {Quay}},
	\bibinfo {author} {\bibfnamefont {P.}~\bibnamefont {Morfin}}, \bibinfo
	{author} {\bibfnamefont {C.}~\bibnamefont {Bena}}, \bibinfo {author}
	{\bibfnamefont {A.~L.}\ \bibnamefont {Yeyati}}, \ and\ \bibinfo {author}
	{\bibfnamefont {P.}~\bibnamefont {Joyez}},\ }\href
{http://dx.doi.org/10.1038/nphys1811} {\bibfield  {journal} {\bibinfo
		{journal} {Nat. Phys.}\ }\textbf {\bibinfo {volume} {6}},\ \bibinfo {pages}
	{965} (\bibinfo {year} {2010})}\BibitemShut {NoStop}%
\bibitem [{\citenamefont {Maurand}\ \emph {et~al.}(2012)\citenamefont
	{Maurand}, \citenamefont {Meng}, \citenamefont {Bonet}, \citenamefont
	{Florens}, \citenamefont {Marty},\ and\ \citenamefont
	{Wernsdorfer}}]{Maurand-2012}%
\BibitemOpen
\bibfield  {author} {\bibinfo {author} {\bibfnamefont {R.}~\bibnamefont
		{Maurand}}, \bibinfo {author} {\bibfnamefont {T.}~\bibnamefont {Meng}},
	\bibinfo {author} {\bibfnamefont {E.}~\bibnamefont {Bonet}}, \bibinfo
	{author} {\bibfnamefont {S.}~\bibnamefont {Florens}}, \bibinfo {author}
	{\bibfnamefont {L.}~\bibnamefont {Marty}}, \ and\ \bibinfo {author}
	{\bibfnamefont {W.}~\bibnamefont {Wernsdorfer}},\ }\href
{http://link.aps.org/doi/10.1103/PhysRevX.2.011009} {\bibfield  {journal}
	{\bibinfo  {journal} {Phys. Rev. X}\ }\textbf {\bibinfo {volume} {2}},\
	\bibinfo {pages} {011009} (\bibinfo {year} {2012})}\BibitemShut {NoStop}%
\bibitem [{\citenamefont {Pillet}\ \emph {et~al.}(2013)\citenamefont {Pillet},
	\citenamefont {Joyez}, \citenamefont {\v{Z}itko},\ and\ \citenamefont
	{Goffman}}]{Pillet-2013}%
\BibitemOpen
\bibfield  {author} {\bibinfo {author} {\bibfnamefont {J.~D.}\ \bibnamefont
		{Pillet}}, \bibinfo {author} {\bibfnamefont {P.}~\bibnamefont {Joyez}},
	\bibinfo {author} {\bibfnamefont {R.}~\bibnamefont {\v{Z}itko}}, \ and\
	\bibinfo {author} {\bibfnamefont {M.~F.}\ \bibnamefont {Goffman}},\ }\href
{http://link.aps.org/doi/10.1103/PhysRevB.88.045101} {\bibfield  {journal}
	{\bibinfo  {journal} {Phys. Rev. B}\ }\textbf {\bibinfo {volume} {88}},\
	\bibinfo {pages} {045101} (\bibinfo {year} {2013})}\BibitemShut {NoStop}%
\bibitem [{\citenamefont {Delagrange}\ \emph {et~al.}(2015)\citenamefont
	{Delagrange}, \citenamefont {Luitz}, \citenamefont {Weil}, \citenamefont
	{Kasumov}, \citenamefont {Meden}, \citenamefont {Bouchiat},\ and\
	\citenamefont {Deblock}}]{Delagrange-2015}%
\BibitemOpen
\bibfield  {author} {\bibinfo {author} {\bibfnamefont {R.}~\bibnamefont
		{Delagrange}}, \bibinfo {author} {\bibfnamefont {D.~J.}\ \bibnamefont
		{Luitz}}, \bibinfo {author} {\bibfnamefont {R.}~\bibnamefont {Weil}},
	\bibinfo {author} {\bibfnamefont {A.}~\bibnamefont {Kasumov}}, \bibinfo
	{author} {\bibfnamefont {V.}~\bibnamefont {Meden}}, \bibinfo {author}
	{\bibfnamefont {H.}~\bibnamefont {Bouchiat}}, \ and\ \bibinfo {author}
	{\bibfnamefont {R.}~\bibnamefont {Deblock}},\ }\href {\doibase
	10.1103/PhysRevB.91.241401} {\bibfield  {journal} {\bibinfo  {journal} {Phys.
			Rev. B}\ }\textbf {\bibinfo {volume} {91}},\ \bibinfo {pages} {241401(R)}
	(\bibinfo {year} {2015})}\BibitemShut {NoStop}%
\bibitem [{\citenamefont {van Dam}\ \emph {et~al.}(2006)\citenamefont {van
		Dam}, \citenamefont {Nazarov}, \citenamefont {Bakkers}, \citenamefont
	{De~Franceschi},\ and\ \citenamefont {Kouwenhoven}}]{vanDam-2006}%
\BibitemOpen
\bibfield  {author} {\bibinfo {author} {\bibfnamefont {J.~A.}\ \bibnamefont
		{van Dam}}, \bibinfo {author} {\bibfnamefont {Y.~V.}\ \bibnamefont
		{Nazarov}}, \bibinfo {author} {\bibfnamefont {E.~P. A.~M.}\ \bibnamefont
		{Bakkers}}, \bibinfo {author} {\bibfnamefont {S.}~\bibnamefont
		{De~Franceschi}}, \ and\ \bibinfo {author} {\bibfnamefont {L.~P.}\
		\bibnamefont {Kouwenhoven}},\ }\href {http://dx.doi.org/10.1038/nature05018}
{\bibfield  {journal} {\bibinfo  {journal} {Nature}\ }\textbf {\bibinfo
		{volume} {442}},\ \bibinfo {pages} {667} (\bibinfo {year}
	{2006})}\BibitemShut {NoStop}%
\bibitem [{\citenamefont {Lee}\ \emph {et~al.}(2012)\citenamefont {Lee},
	\citenamefont {Jiang}, \citenamefont {Aguado}, \citenamefont {Katsaros},
	\citenamefont {Lieber},\ and\ \citenamefont {De~Franceschi}}]{Lee-2012}%
\BibitemOpen
\bibfield  {author} {\bibinfo {author} {\bibfnamefont {E.~J.~H.}\
		\bibnamefont {Lee}}, \bibinfo {author} {\bibfnamefont {X.}~\bibnamefont
		{Jiang}}, \bibinfo {author} {\bibfnamefont {R.}~\bibnamefont {Aguado}},
	\bibinfo {author} {\bibfnamefont {G.}~\bibnamefont {Katsaros}}, \bibinfo
	{author} {\bibfnamefont {C.~M.}\ \bibnamefont {Lieber}}, \ and\ \bibinfo
	{author} {\bibfnamefont {S.}~\bibnamefont {De~Franceschi}},\ }\href {\doibase
	10.1103/PhysRevLett.109.186802} {\bibfield  {journal} {\bibinfo  {journal}
		{Phys. Rev. Lett.}\ }\textbf {\bibinfo {volume} {109}},\ \bibinfo {pages}
	{186802} (\bibinfo {year} {2012})}\BibitemShut {NoStop}%
\bibitem [{\citenamefont {Lee}\ \emph {et~al.}(2017)\citenamefont {Lee},
	\citenamefont {Jiang}, \citenamefont {\ifmmode~\check{Z}\else
		\v{Z}\fi{}itko}, \citenamefont {Aguado}, \citenamefont {Lieber},\ and\
	\citenamefont {De~Franceschi}}]{Lee-2016}%
\BibitemOpen
\bibfield  {author} {\bibinfo {author} {\bibfnamefont {E.~J.~H.}\
		\bibnamefont {Lee}}, \bibinfo {author} {\bibfnamefont {X.}~\bibnamefont
		{Jiang}}, \bibinfo {author} {\bibfnamefont {R.}~\bibnamefont
		{\ifmmode~\check{Z}\else \v{Z}\fi{}itko}}, \bibinfo {author} {\bibfnamefont
		{R.}~\bibnamefont {Aguado}}, \bibinfo {author} {\bibfnamefont {C.~M.}\
		\bibnamefont {Lieber}}, \ and\ \bibinfo {author} {\bibfnamefont
		{S.}~\bibnamefont {De~Franceschi}},\ }\href {\doibase
	10.1103/PhysRevB.95.180502} {\bibfield  {journal} {\bibinfo  {journal} {Phys.
			Rev. B}\ }\textbf {\bibinfo {volume} {95}},\ \bibinfo {pages} {180502}
	(\bibinfo {year} {2017})}\BibitemShut {NoStop}%
\bibitem [{\citenamefont {Li}\ \emph {et~al.}(2017)\citenamefont {Li},
	\citenamefont {Kang}, \citenamefont {Caroff},\ and\ \citenamefont
	{Xu}}]{Xu-2017}%
\BibitemOpen
\bibfield  {author} {\bibinfo {author} {\bibfnamefont {S.}~\bibnamefont
		{Li}}, \bibinfo {author} {\bibfnamefont {N.}~\bibnamefont {Kang}}, \bibinfo
	{author} {\bibfnamefont {P.}~\bibnamefont {Caroff}}, \ and\ \bibinfo {author}
	{\bibfnamefont {H.~Q.}\ \bibnamefont {Xu}},\ }\href {\doibase
	10.1103/PhysRevB.95.014515} {\bibfield  {journal} {\bibinfo  {journal} {Phys.
			Rev. B}\ }\textbf {\bibinfo {volume} {95}},\ \bibinfo {pages} {014515}
	(\bibinfo {year} {2017})}\BibitemShut {NoStop}%
\bibitem [{\citenamefont {Hewson}(1993)}]{Hewson-1993}%
\BibitemOpen
\bibfield  {author} {\bibinfo {author} {\bibfnamefont {A.~C.}\ \bibnamefont
		{Hewson}},\ }\href {\doibase 10.1017/CBO9780511470752} {\emph {\bibinfo
		{title} {The Kondo Problem to Heavy Fermions}}},\ Cambridge Studies in
Magnetism\ (\bibinfo  {publisher} {Cambridge University Press},\ \bibinfo
{year} {1993})\BibitemShut {NoStop}%
\bibitem [{\citenamefont {Wilson}(1975)}]{Wilson-1975}%
\BibitemOpen
\bibfield  {author} {\bibinfo {author} {\bibfnamefont {K.~G.}\ \bibnamefont
		{Wilson}},\ }\href {\doibase 10.1103/RevModPhys.47.773} {\bibfield  {journal}
	{\bibinfo  {journal} {Rev. Mod. Phys.}\ }\textbf {\bibinfo {volume} {47}},\
	\bibinfo {pages} {773} (\bibinfo {year} {1975})}\BibitemShut {NoStop}%
\bibitem [{\citenamefont {Kopietz}\ \emph {et~al.}(2010)\citenamefont
	{Kopietz}, \citenamefont {Bartosch},\ and\ \citenamefont
	{Sch\"{u}tz}}]{Kopietz-2010}%
\BibitemOpen
\bibfield  {author} {\bibinfo {author} {\bibfnamefont {P.}~\bibnamefont
		{Kopietz}}, \bibinfo {author} {\bibfnamefont {L.}~\bibnamefont {Bartosch}}, \
	and\ \bibinfo {author} {\bibfnamefont {F.}~\bibnamefont {Sch\"{u}tz}},\
}\href {\doibase DOI={10.1017/CBO9780511470752},} {\emph {\bibinfo {title}
	{Introduction to the functional renormalization group}}},\ Lect. Notes Phys.\
(\bibinfo  {publisher} {Springer-Verlag Berlin Heidelberg},\ \bibinfo {year}
{2010})\BibitemShut {NoStop}%
\bibitem [{\citenamefont {Streib}\ \emph {et~al.}(2013)\citenamefont {Streib},
	\citenamefont {Isidori},\ and\ \citenamefont {Kopietz}}]{Kopietz-2013}%
\BibitemOpen
\bibfield  {author} {\bibinfo {author} {\bibfnamefont {S.}~\bibnamefont
		{Streib}}, \bibinfo {author} {\bibfnamefont {A.}~\bibnamefont {Isidori}}, \
	and\ \bibinfo {author} {\bibfnamefont {P.}~\bibnamefont {Kopietz}},\ }\href
{\doibase 10.1103/PhysRevB.87.201107} {\bibfield  {journal} {\bibinfo
		{journal} {Phys. Rev. B}\ }\textbf {\bibinfo {volume} {87}},\ \bibinfo
	{pages} {201107} (\bibinfo {year} {2013})}\BibitemShut {NoStop}%
\bibitem [{\citenamefont {Edwards}\ \emph {et~al.}(2013)\citenamefont
	{Edwards}, \citenamefont {Hewson},\ and\ \citenamefont
	{Pandis}}]{Hewson-RG-2013}%
\BibitemOpen
\bibfield  {author} {\bibinfo {author} {\bibfnamefont {K.}~\bibnamefont
		{Edwards}}, \bibinfo {author} {\bibfnamefont {A.~C.}\ \bibnamefont {Hewson}},
	\ and\ \bibinfo {author} {\bibfnamefont {V.}~\bibnamefont {Pandis}},\ }\href
{\doibase 10.1103/PhysRevB.87.165128} {\bibfield  {journal} {\bibinfo
		{journal} {Phys. Rev. B}\ }\textbf {\bibinfo {volume} {87}},\ \bibinfo
	{pages} {165128} (\bibinfo {year} {2013})}\BibitemShut {NoStop}%
\bibitem [{\citenamefont {Jani{\v s}}\ and\ \citenamefont
	{Augustinsk{\'y}}(2007)}]{Janis-2007}%
\BibitemOpen
\bibfield  {author} {\bibinfo {author} {\bibfnamefont {V.}~\bibnamefont
		{Jani{\v s}}}\ and\ \bibinfo {author} {\bibfnamefont {P.}~\bibnamefont
		{Augustinsk{\'y}}},\ }\href
{http://link.aps.org/doi/10.1103/PhysRevB.75.165108} {\bibfield  {journal}
	{\bibinfo  {journal} {Phys. Rev. B}\ }\textbf {\bibinfo {volume} {75}},\
	\bibinfo {pages} {165108} (\bibinfo {year} {2007})}\BibitemShut {NoStop}%
\bibitem [{\citenamefont {Buitelaar}\ \emph {et~al.}(2002)\citenamefont
	{Buitelaar}, \citenamefont {Nussbaumer},\ and\ \citenamefont
	{Sch{\"o}nenberger}}]{Buitelaar-2002}%
\BibitemOpen
\bibfield  {author} {\bibinfo {author} {\bibfnamefont {M.~R.}\ \bibnamefont
		{Buitelaar}}, \bibinfo {author} {\bibfnamefont {T.}~\bibnamefont
		{Nussbaumer}}, \ and\ \bibinfo {author} {\bibfnamefont {C.}~\bibnamefont
		{Sch{\"o}nenberger}},\ }\href
{http://link.aps.org/doi/10.1103/PhysRevLett.89.256801} {\bibfield  {journal}
	{\bibinfo  {journal} {Phys. Rev. Lett.}\ }\textbf {\bibinfo {volume} {89}},\
	\bibinfo {pages} {256801} (\bibinfo {year} {2002})}\BibitemShut {NoStop}%
\bibitem [{\citenamefont {Eichler}\ \emph {et~al.}(2007)\citenamefont
	{Eichler}, \citenamefont {Weiss}, \citenamefont {Oberholzer}, \citenamefont
	{Sch\"onenberger}, \citenamefont {Levy~Yeyati}, \citenamefont {Cuevas},\ and\
	\citenamefont {Mart\'{\i}n-Rodero}}]{Eichler-2007}%
\BibitemOpen
\bibfield  {author} {\bibinfo {author} {\bibfnamefont {A.}~\bibnamefont
		{Eichler}}, \bibinfo {author} {\bibfnamefont {M.}~\bibnamefont {Weiss}},
	\bibinfo {author} {\bibfnamefont {S.}~\bibnamefont {Oberholzer}}, \bibinfo
	{author} {\bibfnamefont {C.}~\bibnamefont {Sch\"onenberger}}, \bibinfo
	{author} {\bibfnamefont {A.}~\bibnamefont {Levy~Yeyati}}, \bibinfo {author}
	{\bibfnamefont {J.~C.}\ \bibnamefont {Cuevas}}, \ and\ \bibinfo {author}
	{\bibfnamefont {A.}~\bibnamefont {Mart\'{\i}n-Rodero}},\ }\href {\doibase
	10.1103/PhysRevLett.99.126602} {\bibfield  {journal} {\bibinfo  {journal}
		{Phys. Rev. Lett.}\ }\textbf {\bibinfo {volume} {99}},\ \bibinfo {pages}
	{126602} (\bibinfo {year} {2007})}\BibitemShut {NoStop}%
\bibitem [{\citenamefont {Sand-Jespersen}\ \emph {et~al.}(2007)\citenamefont
	{Sand-Jespersen}, \citenamefont {Paaske}, \citenamefont {Andersen},
	\citenamefont {Grove-Rasmussen}, \citenamefont {J\o{}rgensen}, \citenamefont
	{Aagesen}, \citenamefont {S\o{}rensen}, \citenamefont {Lindelof},
	\citenamefont {Flensberg},\ and\ \citenamefont
	{Nyg\aa{}rd}}]{Sand-Jespersen-2007}%
\BibitemOpen
\bibfield  {author} {\bibinfo {author} {\bibfnamefont {T.}~\bibnamefont
		{Sand-Jespersen}}, \bibinfo {author} {\bibfnamefont {J.}~\bibnamefont
		{Paaske}}, \bibinfo {author} {\bibfnamefont {B.~M.}\ \bibnamefont
		{Andersen}}, \bibinfo {author} {\bibfnamefont {K.}~\bibnamefont
		{Grove-Rasmussen}}, \bibinfo {author} {\bibfnamefont {H.~I.}\ \bibnamefont
		{J\o{}rgensen}}, \bibinfo {author} {\bibfnamefont {M.}~\bibnamefont
		{Aagesen}}, \bibinfo {author} {\bibfnamefont {C.~B.}\ \bibnamefont
		{S\o{}rensen}}, \bibinfo {author} {\bibfnamefont {P.~E.}\ \bibnamefont
		{Lindelof}}, \bibinfo {author} {\bibfnamefont {K.}~\bibnamefont {Flensberg}},
	\ and\ \bibinfo {author} {\bibfnamefont {J.}~\bibnamefont {Nyg\aa{}rd}},\
}\href {\doibase 10.1103/PhysRevLett.99.126603} {\bibfield  {journal}
{\bibinfo  {journal} {Phys. Rev. Lett.}\ }\textbf {\bibinfo {volume} {99}},\
\bibinfo {pages} {126603} (\bibinfo {year} {2007})}\BibitemShut {NoStop}%
\bibitem [{\citenamefont {Buizert}\ \emph {et~al.}(2007)\citenamefont
	{Buizert}, \citenamefont {Oiwa}, \citenamefont {Shibata}, \citenamefont
	{Hirakawa},\ and\ \citenamefont {Tarucha}}]{Buizert-2007}%
\BibitemOpen
\bibfield  {author} {\bibinfo {author} {\bibfnamefont {C.}~\bibnamefont
		{Buizert}}, \bibinfo {author} {\bibfnamefont {A.}~\bibnamefont {Oiwa}},
	\bibinfo {author} {\bibfnamefont {K.}~\bibnamefont {Shibata}}, \bibinfo
	{author} {\bibfnamefont {K.}~\bibnamefont {Hirakawa}}, \ and\ \bibinfo
	{author} {\bibfnamefont {S.}~\bibnamefont {Tarucha}},\ }\href {\doibase
	10.1103/PhysRevLett.99.136806} {\bibfield  {journal} {\bibinfo  {journal}
		{Phys. Rev. Lett.}\ }\textbf {\bibinfo {volume} {99}},\ \bibinfo {pages}
	{136806} (\bibinfo {year} {2007})}\BibitemShut {NoStop}%
\bibitem [{\citenamefont {Grove-Rasmussen}\ \emph {et~al.}(2007)\citenamefont
	{Grove-Rasmussen}, \citenamefont {J{\o}rgensen},\ and\ \citenamefont
	{Lindelof}}]{Grove-2007}%
\BibitemOpen
\bibfield  {author} {\bibinfo {author} {\bibfnamefont {K.}~\bibnamefont
		{Grove-Rasmussen}}, \bibinfo {author} {\bibfnamefont {H.~I.}\ \bibnamefont
		{J{\o}rgensen}}, \ and\ \bibinfo {author} {\bibfnamefont {P.~E.}\
		\bibnamefont {Lindelof}},\ }\href
{http://stacks.iop.org/1367-2630/9/i=5/a=124} {\bibfield  {journal} {\bibinfo
		{journal} {New J. Phys.}\ }\textbf {\bibinfo {volume} {9}},\ \bibinfo
	{pages} {124} (\bibinfo {year} {2007})}\BibitemShut {NoStop}%
\bibitem [{\citenamefont {Luitz}\ \emph {et~al.}(2012)\citenamefont {Luitz},
	\citenamefont {Assaad}, \citenamefont {Novotn{\'y}}, \citenamefont
	{Karrasch},\ and\ \citenamefont {Meden}}]{Luitz-2012}%
\BibitemOpen
\bibfield  {author} {\bibinfo {author} {\bibfnamefont {D.~J.}\ \bibnamefont
		{Luitz}}, \bibinfo {author} {\bibfnamefont {F.~F.}\ \bibnamefont {Assaad}},
	\bibinfo {author} {\bibfnamefont {T.}~\bibnamefont {Novotn{\'y}}}, \bibinfo
	{author} {\bibfnamefont {C.}~\bibnamefont {Karrasch}}, \ and\ \bibinfo
	{author} {\bibfnamefont {V.}~\bibnamefont {Meden}},\ }\href
{http://link.aps.org/doi/10.1103/PhysRevLett.108.227001} {\bibfield
	{journal} {\bibinfo  {journal} {Phys. Rev. Lett.}\ }\textbf {\bibinfo
		{volume} {108}},\ \bibinfo {pages} {227001} (\bibinfo {year}
	{2012})}\BibitemShut {NoStop}%
\bibitem [{\citenamefont {Mart{\'\i}n-Rodero}\ and\ \citenamefont
	{Levy~Yeyati}(2011)}]{Rodero-2011}%
\BibitemOpen
\bibfield  {author} {\bibinfo {author} {\bibfnamefont {A.}~\bibnamefont
		{Mart{\'\i}n-Rodero}}\ and\ \bibinfo {author} {\bibfnamefont
		{A.}~\bibnamefont {Levy~Yeyati}},\ }\href {\doibase
	10.1080/00018732.2011.624266} {\bibfield  {journal} {\bibinfo  {journal}
		{Adv. Phys.}\ }\textbf {\bibinfo {volume} {60}},\ \bibinfo {pages} {899}
	(\bibinfo {year} {2011})}\BibitemShut {NoStop}%
\bibitem [{\citenamefont {Meden}(2019)}]{Meden-2019}%
\BibitemOpen
\bibfield  {author} {\bibinfo {author} {\bibfnamefont {V.}~\bibnamefont
		{Meden}},\ }\href {\doibase 10.1088/1361-648x/aafd6a} {\bibfield  {journal}
	{\bibinfo  {journal} {Journal of Physics: Condensed Matter}\ }\textbf
	{\bibinfo {volume} {31}},\ \bibinfo {pages} {163001} (\bibinfo {year}
	{2019})}\BibitemShut {NoStop}%
\bibitem [{\citenamefont {De~Franceschi}\ \emph {et~al.}(2010)\citenamefont
	{De~Franceschi}, \citenamefont {Kouwenhoven}, \citenamefont
	{Sch{\"o}nenberger},\ and\ \citenamefont {Wernsdorfer}}]{Wernsdorfer-2010}%
\BibitemOpen
\bibfield  {author} {\bibinfo {author} {\bibfnamefont {S.}~\bibnamefont
		{De~Franceschi}}, \bibinfo {author} {\bibfnamefont {L.}~\bibnamefont
		{Kouwenhoven}}, \bibinfo {author} {\bibfnamefont {C.}~\bibnamefont
		{Sch{\"o}nenberger}}, \ and\ \bibinfo {author} {\bibfnamefont
		{W.}~\bibnamefont {Wernsdorfer}},\ }\href
{http://dx.doi.org/10.1038/nnano.2010.173} {\bibfield  {journal} {\bibinfo
		{journal} {Nat. Nanotechnol.}\ }\textbf {\bibinfo {volume} {5}},\ \bibinfo
	{pages} {703} (\bibinfo {year} {2010})}\BibitemShut {NoStop}%
\bibitem [{\citenamefont {J{\o}rgensen}\ \emph {et~al.}(2007)\citenamefont
	{J{\o}rgensen}, \citenamefont {Novotn{\'y}}, \citenamefont {Grove-Rasmussen},
	\citenamefont {Flensberg},\ and\ \citenamefont {Lindelof}}]{Jorgensen-2007}%
\BibitemOpen
\bibfield  {author} {\bibinfo {author} {\bibfnamefont {H.~I.}\ \bibnamefont
		{J{\o}rgensen}}, \bibinfo {author} {\bibfnamefont {T.}~\bibnamefont
		{Novotn{\'y}}}, \bibinfo {author} {\bibfnamefont {K.}~\bibnamefont
		{Grove-Rasmussen}}, \bibinfo {author} {\bibfnamefont {K.}~\bibnamefont
		{Flensberg}}, \ and\ \bibinfo {author} {\bibfnamefont {P.~E.}\ \bibnamefont
		{Lindelof}},\ }\href {\doibase 10.1021/nl071152w} {\bibfield  {journal}
	{\bibinfo  {journal} {Nano Lett.}\ }\textbf {\bibinfo {volume} {7}},\
	\bibinfo {pages} {2441} (\bibinfo {year} {2007})}\BibitemShut {NoStop}%
\bibitem [{\citenamefont {{\v Z}itko}\ \emph {et~al.}(2015)\citenamefont {{\v
			Z}itko}, \citenamefont {Lim}, \citenamefont {L{\'o}pez},\ and\ \citenamefont
	{Aguado}}]{Zitko-2015}%
\BibitemOpen
\bibfield  {author} {\bibinfo {author} {\bibfnamefont {R.}~\bibnamefont {{\v
				Z}itko}}, \bibinfo {author} {\bibfnamefont {J.~S.}\ \bibnamefont {Lim}},
	\bibinfo {author} {\bibfnamefont {R.}~\bibnamefont {L{\'o}pez}}, \ and\
	\bibinfo {author} {\bibfnamefont {R.}~\bibnamefont {Aguado}},\ }\href
{http://link.aps.org/doi/10.1103/PhysRevB.91.045441} {\bibfield  {journal}
	{\bibinfo  {journal} {Phys. Rev. B}\ }\textbf {\bibinfo {volume} {91}},\
	\bibinfo {pages} {045441} (\bibinfo {year} {2015})}\BibitemShut {NoStop}%
\bibitem [{\citenamefont {Kir\ifmmode~\check{s}\else \v{s}\fi{}anskas}\ \emph
	{et~al.}(2015)\citenamefont {Kir\ifmmode~\check{s}\else \v{s}\fi{}anskas},
	\citenamefont {Goldstein}, \citenamefont {Flensberg}, \citenamefont
	{Glazman},\ and\ \citenamefont {Paaske}}]{Paaske-2015}%
\BibitemOpen
\bibfield  {author} {\bibinfo {author} {\bibfnamefont {G.}~\bibnamefont
		{Kir\ifmmode~\check{s}\else \v{s}\fi{}anskas}}, \bibinfo {author}
	{\bibfnamefont {M.}~\bibnamefont {Goldstein}}, \bibinfo {author}
	{\bibfnamefont {K.}~\bibnamefont {Flensberg}}, \bibinfo {author}
	{\bibfnamefont {L.~I.}\ \bibnamefont {Glazman}}, \ and\ \bibinfo {author}
	{\bibfnamefont {J.}~\bibnamefont {Paaske}},\ }\href {\doibase
	10.1103/PhysRevB.92.235422} {\bibfield  {journal} {\bibinfo  {journal} {Phys.
			Rev. B}\ }\textbf {\bibinfo {volume} {92}},\ \bibinfo {pages} {235422}
	(\bibinfo {year} {2015})}\BibitemShut {NoStop}%
\bibitem [{\citenamefont {Jellinggaard}\ \emph {et~al.}(2016)\citenamefont
	{Jellinggaard}, \citenamefont {Grove-Rasmussen}, \citenamefont {Madsen},\
	and\ \citenamefont {Nyg{\aa}rd}}]{Jellinggaard-2016}%
\BibitemOpen
\bibfield  {author} {\bibinfo {author} {\bibfnamefont {A.}~\bibnamefont
		{Jellinggaard}}, \bibinfo {author} {\bibfnamefont {K.}~\bibnamefont
		{Grove-Rasmussen}}, \bibinfo {author} {\bibfnamefont {M.~H.}\ \bibnamefont
		{Madsen}}, \ and\ \bibinfo {author} {\bibfnamefont {J.}~\bibnamefont
		{Nyg{\aa}rd}},\ }\href {http://link.aps.org/doi/10.1103/PhysRevB.94.064520}
{\bibfield  {journal} {\bibinfo  {journal} {Phys. Rev. B}\ }\textbf {\bibinfo
		{volume} {94}},\ \bibinfo {pages} {064520} (\bibinfo {year}
	{2016})}\BibitemShut {NoStop}%
\bibitem [{\citenamefont {Doma{\'n}ski}\ \emph {et~al.}(2017)\citenamefont
	{Doma{\'n}ski}, \citenamefont {{\v Z}onda}, \citenamefont {Pokorn{\'y}},
	\citenamefont {G{\'o}rski}, \citenamefont {Jani{\v s}},\ and\ \citenamefont
	{Novotn{\'y}}}]{Domanski-2017}%
\BibitemOpen
\bibfield  {author} {\bibinfo {author} {\bibfnamefont {T.}~\bibnamefont
		{Doma{\'n}ski}}, \bibinfo {author} {\bibfnamefont {M.}~\bibnamefont {{\v
				Z}onda}}, \bibinfo {author} {\bibfnamefont {V.}~\bibnamefont {Pokorn{\'y}}},
	\bibinfo {author} {\bibfnamefont {G.}~\bibnamefont {G{\'o}rski}}, \bibinfo
	{author} {\bibfnamefont {V.}~\bibnamefont {Jani{\v s}}}, \ and\ \bibinfo
	{author} {\bibfnamefont {T.}~\bibnamefont {Novotn{\'y}}},\ }\href
{http://link.aps.org/doi/10.1103/PhysRevB.95.045104} {\bibfield  {journal}
	{\bibinfo  {journal} {Physical Review B}\ }\textbf {\bibinfo {volume} {95}},\
	\bibinfo {pages} {045104} (\bibinfo {year} {2017})}\BibitemShut {NoStop}%
\bibitem [{\citenamefont {Satori}\ \emph {et~al.}(1992)\citenamefont {Satori},
	\citenamefont {Shiba}, \citenamefont {Sakai},\ and\ \citenamefont
	{Shimizu}}]{Satori-1992}%
\BibitemOpen
\bibfield  {author} {\bibinfo {author} {\bibfnamefont {K.}~\bibnamefont
		{Satori}}, \bibinfo {author} {\bibfnamefont {H.}~\bibnamefont {Shiba}},
	\bibinfo {author} {\bibfnamefont {O.}~\bibnamefont {Sakai}}, \ and\ \bibinfo
	{author} {\bibfnamefont {Y.}~\bibnamefont {Shimizu}},\ }\href {\doibase
	10.1143/JPSJ.61.3239} {\bibfield  {journal} {\bibinfo  {journal} {J. Phys.
			Soc. Japan.}\ }\textbf {\bibinfo {volume} {61}},\ \bibinfo {pages} {3239}
	(\bibinfo {year} {1992})}\BibitemShut {NoStop}%
\bibitem [{\citenamefont {Yoshioka}\ and\ \citenamefont
	{Ohashi}(2000)}]{Yoshioka-2000}%
\BibitemOpen
\bibfield  {author} {\bibinfo {author} {\bibfnamefont {T.}~\bibnamefont
		{Yoshioka}}\ and\ \bibinfo {author} {\bibfnamefont {Y.}~\bibnamefont
		{Ohashi}},\ }\href {\doibase 10.1143/JPSJ.69.1812} {\bibfield  {journal}
	{\bibinfo  {journal} {J. Phys. Soc. Jpn.}\ }\textbf {\bibinfo {volume}
		{69}},\ \bibinfo {pages} {1812} (\bibinfo {year} {2000})}\BibitemShut
{NoStop}%
\bibitem [{\citenamefont {Tanaka}\ \emph {et~al.}(2007)\citenamefont {Tanaka},
	\citenamefont {Oguri},\ and\ \citenamefont {Hewson}}]{Tanaka-2007}%
\BibitemOpen
\bibfield  {author} {\bibinfo {author} {\bibfnamefont {Y.}~\bibnamefont
		{Tanaka}}, \bibinfo {author} {\bibfnamefont {A.}~\bibnamefont {Oguri}}, \
	and\ \bibinfo {author} {\bibfnamefont {A.~C.}\ \bibnamefont {Hewson}},\
}\href {http://stacks.iop.org/1367-2630/9/i=5/a=115} {\bibfield  {journal}
{\bibinfo  {journal} {New J. Phys.}\ }\textbf {\bibinfo {volume} {9}},\
\bibinfo {pages} {115} (\bibinfo {year} {2007})}\BibitemShut {NoStop}%
\bibitem [{\citenamefont {Oguri}\ and\ \citenamefont
	{Tanaka}(2012)}]{Oguri-2012}%
\BibitemOpen
\bibfield  {author} {\bibinfo {author} {\bibfnamefont {A.}~\bibnamefont
		{Oguri}}\ and\ \bibinfo {author} {\bibfnamefont {Y.}~\bibnamefont {Tanaka}},\
}\href {http://stacks.iop.org/1742-6596/391/i=1/a=012146} {\bibfield
{journal} {\bibinfo  {journal} {J. Phys.: Conf. Ser.}\ }\textbf {\bibinfo
	{volume} {391}},\ \bibinfo {pages} {012146} (\bibinfo {year}
{2012})}\BibitemShut {NoStop}%
\bibitem [{\citenamefont {Oguri}\ \emph {et~al.}(2013)\citenamefont {Oguri},
	\citenamefont {Tanaka},\ and\ \citenamefont {Bauer}}]{Oguri-2013}%
\BibitemOpen
\bibfield  {author} {\bibinfo {author} {\bibfnamefont {A.}~\bibnamefont
		{Oguri}}, \bibinfo {author} {\bibfnamefont {Y.}~\bibnamefont {Tanaka}}, \
	and\ \bibinfo {author} {\bibfnamefont {J.}~\bibnamefont {Bauer}},\ }\href
{\doibase 10.1103/PhysRevB.87.075432} {\bibfield  {journal} {\bibinfo
		{journal} {Phys. Rev. B}\ }\textbf {\bibinfo {volume} {87}},\ \bibinfo
	{pages} {075432} (\bibinfo {year} {2013})}\BibitemShut {NoStop}%
\bibitem [{\citenamefont {Liu}\ \emph {et~al.}(2016)\citenamefont {Liu},
	\citenamefont {Wang},\ and\ \citenamefont {Wang}}]{Liu-2016}%
\BibitemOpen
\bibfield  {author} {\bibinfo {author} {\bibfnamefont {J.-G.}\ \bibnamefont
		{Liu}}, \bibinfo {author} {\bibfnamefont {D.}~\bibnamefont {Wang}}, \ and\
	\bibinfo {author} {\bibfnamefont {Q.-H.}\ \bibnamefont {Wang}},\ }\href
{\doibase 10.1103/PhysRevB.93.035102} {\bibfield  {journal} {\bibinfo
		{journal} {Phys. Rev. B}\ }\textbf {\bibinfo {volume} {93}},\ \bibinfo
	{pages} {035102} (\bibinfo {year} {2016})}\BibitemShut {NoStop}%
\bibitem [{Note1()}]{Note1}%
\BibitemOpen
\bibinfo {note} {In the present model, half-filling implies that the system
	is tuned to the particle-hole symmetric point, see Ref.~\cite
	{Tanaka-2007}.}\BibitemShut {Stop}%
\bibitem [{\citenamefont {Bulla}\ \emph {et~al.}(1994)\citenamefont {Bulla},
	\citenamefont {Keller},\ and\ \citenamefont {Pruschke}}]{Bulla-1994}%
\BibitemOpen
\bibfield  {author} {\bibinfo {author} {\bibfnamefont {R.}~\bibnamefont
		{Bulla}}, \bibinfo {author} {\bibfnamefont {J.}~\bibnamefont {Keller}}, \
	and\ \bibinfo {author} {\bibfnamefont {T.}~\bibnamefont {Pruschke}},\ }\href
{\doibase 10.1007/BF01307671} {\bibfield  {journal} {\bibinfo  {journal} {Z.
			Phys. B Condens. Matter}\ }\textbf {\bibinfo {volume} {94}},\ \bibinfo
	{pages} {195} (\bibinfo {year} {1994})}\BibitemShut {NoStop}%
\bibitem [{\citenamefont {Hecht}\ \emph {et~al.}(2008)\citenamefont {Hecht},
	\citenamefont {Weichselbaum}, \citenamefont {von Delft},\ and\ \citenamefont
	{Bulla}}]{Hecht-2008}%
\BibitemOpen
\bibfield  {author} {\bibinfo {author} {\bibfnamefont {T.}~\bibnamefont
		{Hecht}}, \bibinfo {author} {\bibfnamefont {A.}~\bibnamefont {Weichselbaum}},
	\bibinfo {author} {\bibfnamefont {J.}~\bibnamefont {von Delft}}, \ and\
	\bibinfo {author} {\bibfnamefont {R.}~\bibnamefont {Bulla}},\ }\href
{http://stacks.iop.org/0953-8984/20/i=27/a=275213} {\bibfield  {journal}
	{\bibinfo  {journal} {J. Phys.: Cond. Mat.}\ }\textbf {\bibinfo {volume}
		{20}},\ \bibinfo {pages} {275213} (\bibinfo {year} {2008})}\BibitemShut
{NoStop}%
\bibitem [{\citenamefont {{\v Z}itko}(2014)}]{Ljubljana-code}%
\BibitemOpen
\bibfield  {author} {\bibinfo {author} {\bibfnamefont {R.}~\bibnamefont {{\v
				Z}itko}},\ }\href@noop {} {\enquote {\bibinfo {title} {{NRG L}jubljana - open
			source numerical renormalization group code},}\ } (\bibinfo {year} {2014}),\
\bibinfo {note} {nrgljubljana.ijs.si}\BibitemShut {NoStop}%
\bibitem [{\citenamefont {Kadlecov{\'a}}\ \emph {et~al.}(2017)\citenamefont
	{Kadlecov{\'a}}, \citenamefont {{\v Z}onda},\ and\ \citenamefont
	{Novotn{\'y}}}]{Kadlecova-2017}%
\BibitemOpen
\bibfield  {author} {\bibinfo {author} {\bibfnamefont {A.}~\bibnamefont
		{Kadlecov{\'a}}}, \bibinfo {author} {\bibfnamefont {M.}~\bibnamefont {{\v
				Z}onda}}, \ and\ \bibinfo {author} {\bibfnamefont {T.}~\bibnamefont
		{Novotn{\'y}}},\ }\href {\doibase 10.1103/PhysRevB.95.195114} {\bibfield
	{journal} {\bibinfo  {journal} {Phys. Rev. B}\ }\textbf {\bibinfo {volume}
		{95}},\ \bibinfo {pages} {195114} (\bibinfo {year} {2017})}\BibitemShut
{NoStop}%
\bibitem [{Note2()}]{Note2}%
\BibitemOpen
\bibinfo {note} {The argument $\omega ^+$ of the function emphasizes to which
	part of the complex $z$-plain it belongs (above vs. below the real axis of
	$z$). This notation is employed throughout this paper when
	required.}\BibitemShut {Stop}%
\bibitem [{\citenamefont {Novotn\'y}\ \emph {et~al.}(2005)\citenamefont
	{Novotn\'y}, \citenamefont {Rossini},\ and\ \citenamefont
	{Flensberg}}]{Novotny-Rossini-2005}%
\BibitemOpen
\bibfield  {author} {\bibinfo {author} {\bibfnamefont {T.}~\bibnamefont
		{Novotn\'y}}, \bibinfo {author} {\bibfnamefont {A.}~\bibnamefont {Rossini}},
	\ and\ \bibinfo {author} {\bibfnamefont {K.}~\bibnamefont {Flensberg}},\
}\href {\doibase 10.1103/PhysRevB.72.224502} {\bibfield  {journal} {\bibinfo
	{journal} {Phys. Rev. B}\ }\textbf {\bibinfo {volume} {72}},\ \bibinfo
{pages} {224502} (\bibinfo {year} {2005})}\BibitemShut {NoStop}%
\bibitem [{Note3()}]{Note3}%
\BibitemOpen
\bibinfo {note} {The spectral functions have been obtained using the open
	source NRG Ljubljana code \cite {Ljubljana-code} in the one-channel mode with
	intertwined $z$-discretization \cite {ZitkoPruschke-2009} with $z= n/10$, $n
	\in \protect \{ 0, \protect \ldots 10 \protect \}$}\BibitemShut {NoStop}%
\bibitem [{Note4()}]{Note4}%
\BibitemOpen
\bibinfo {note} {The broadening has been also predicted analytically in
	Refs.~\cite {Domanski-2016,Domanski-2017} via the Schrieffer-Wolff
	transformation. Using the transformation $\protect \mathbb {T}$, the $\Delta
	\rightarrow \infty $ can be mapped onto the particle-hole asymmetric SIAM,
	from which the enhancement of the exchange coupling $J$ compared to the
	particle-hole symmetric case follows trivially. Thus, as a consequence of the
	locally induced SC pairing, the $T_K$ is enhanced and the central peak is
	broader compared to the single-channel normal SIAM.}\BibitemShut {Stop}%
\bibitem [{\citenamefont {Bulla}\ \emph {et~al.}(2008)\citenamefont {Bulla},
	\citenamefont {Costi},\ and\ \citenamefont {Pruschke}}]{Bulla-Rev-2008}%
\BibitemOpen
\bibfield  {author} {\bibinfo {author} {\bibfnamefont {R.}~\bibnamefont
		{Bulla}}, \bibinfo {author} {\bibfnamefont {T.~A.}\ \bibnamefont {Costi}}, \
	and\ \bibinfo {author} {\bibfnamefont {T.}~\bibnamefont {Pruschke}},\ }\href
{http://link.aps.org/doi/10.1103/RevModPhys.80.395} {\bibfield  {journal}
	{\bibinfo  {journal} {Rev. Mod. Phys.}\ }\textbf {\bibinfo {volume} {80}},\
	\bibinfo {pages} {395} (\bibinfo {year} {2008})}\BibitemShut {NoStop}%
\bibitem [{\citenamefont {{\v Z}itko}\ and\ \citenamefont
	{Pruschke}(2009)}]{ZitkoPruschke-2009}%
\BibitemOpen
\bibfield  {author} {\bibinfo {author} {\bibfnamefont {R.}~\bibnamefont {{\v
				Z}itko}}\ and\ \bibinfo {author} {\bibfnamefont {T.}~\bibnamefont
		{Pruschke}},\ }\href {\doibase 10.1103/PhysRevB.79.085106} {\bibfield
	{journal} {\bibinfo  {journal} {Phys. Rev. B}\ }\textbf {\bibinfo {volume}
		{79}},\ \bibinfo {pages} {085106} (\bibinfo {year} {2009})}\BibitemShut
{NoStop}%
\bibitem [{\citenamefont {{\v Z}onda}\ \emph {et~al.}(2015)\citenamefont {{\v
			Z}onda}, \citenamefont {Pokorn\'y}, \citenamefont {Jani\v{s}},\ and\
	\citenamefont {Novotn\'y}}]{Zonda-2015}%
\BibitemOpen
\bibfield  {author} {\bibinfo {author} {\bibfnamefont {M.}~\bibnamefont {{\v
				Z}onda}}, \bibinfo {author} {\bibfnamefont {V.}~\bibnamefont {Pokorn\'y}},
	\bibinfo {author} {\bibfnamefont {V.}~\bibnamefont {Jani\v{s}}}, \ and\
	\bibinfo {author} {\bibfnamefont {T.}~\bibnamefont {Novotn\'y}},\ }\href
{\doibase 10.1038/srep08821} {\bibfield  {journal} {\bibinfo  {journal} {Sci.
			Rep.}\ }\textbf {\bibinfo {volume} {5}},\ \bibinfo {pages} {8821} (\bibinfo
	{year} {2015})}\BibitemShut {NoStop}%
\bibitem [{\citenamefont {Doma{\'n}ski}\ \emph {et~al.}(2016)\citenamefont
	{Doma{\'n}ski}, \citenamefont {Weymann}, \citenamefont {Bara{\'n}ska},\ and\
	\citenamefont {G{\'o}rski}}]{Domanski-2016}%
\BibitemOpen
\bibfield  {author} {\bibinfo {author} {\bibfnamefont {T.}~\bibnamefont
		{Doma{\'n}ski}}, \bibinfo {author} {\bibfnamefont {I.}~\bibnamefont
		{Weymann}}, \bibinfo {author} {\bibfnamefont {M.}~\bibnamefont
		{Bara{\'n}ska}}, \ and\ \bibinfo {author} {\bibfnamefont {G.}~\bibnamefont
		{G{\'o}rski}},\ }\href {http://dx.doi.org/10.1038/srep23336} {\bibfield
	{journal} {\bibinfo  {journal} {Sci. Rep.}\ }\textbf {\bibinfo {volume}
		{6}},\ \bibinfo {pages} {23336} (\bibinfo {year} {2016})}\BibitemShut
{NoStop}%
\bibitem [{\citenamefont {Fazio}\ and\ \citenamefont
	{Raimondi}(1998)}]{Fazio-1998}%
\BibitemOpen
\bibfield  {author} {\bibinfo {author} {\bibfnamefont {R.}~\bibnamefont
		{Fazio}}\ and\ \bibinfo {author} {\bibfnamefont {R.}~\bibnamefont
		{Raimondi}},\ }\href {http://link.aps.org/doi/10.1103/PhysRevLett.80.2913}
{\bibfield  {journal} {\bibinfo  {journal} {Phys. Rev. Lett.}\ }\textbf
	{\bibinfo {volume} {80}},\ \bibinfo {pages} {2913} (\bibinfo {year}
	{1998})}\BibitemShut {NoStop}%
\bibitem [{\citenamefont {Mitchell}\ \emph {et~al.}(2014)\citenamefont
	{Mitchell}, \citenamefont {Galpin}, \citenamefont {Wilson-Fletcher},
	\citenamefont {Logan},\ and\ \citenamefont {Bulla}}]{Mitchell-2014}%
\BibitemOpen
\bibfield  {author} {\bibinfo {author} {\bibfnamefont {A.~K.}\ \bibnamefont
		{Mitchell}}, \bibinfo {author} {\bibfnamefont {M.~R.}\ \bibnamefont
		{Galpin}}, \bibinfo {author} {\bibfnamefont {S.}~\bibnamefont
		{Wilson-Fletcher}}, \bibinfo {author} {\bibfnamefont {D.~E.}\ \bibnamefont
		{Logan}}, \ and\ \bibinfo {author} {\bibfnamefont {R.}~\bibnamefont
		{Bulla}},\ }\href {\doibase 10.1103/PhysRevB.89.121105} {\bibfield  {journal}
	{\bibinfo  {journal} {Phys. Rev. B}\ }\textbf {\bibinfo {volume} {89}},\
	\bibinfo {pages} {121105} (\bibinfo {year} {2014})}\BibitemShut {NoStop}%
\bibitem [{\citenamefont {Kadlecov\'a}\ \emph {et~al.}(2019)\citenamefont
	{Kadlecov\'a}, \citenamefont {\ifmmode~\check{Z}\else \v{Z}\fi{}onda},
	\citenamefont {Pokorn\'y},\ and\ \citenamefont {Novotn\'y}}]{Kadlecova-2019}%
\BibitemOpen
\bibfield  {author} {\bibinfo {author} {\bibfnamefont {A.}~\bibnamefont
		{Kadlecov\'a}}, \bibinfo {author} {\bibfnamefont {M.}~\bibnamefont
		{\ifmmode~\check{Z}\else \v{Z}\fi{}onda}}, \bibinfo {author} {\bibfnamefont
		{V.}~\bibnamefont {Pokorn\'y}}, \ and\ \bibinfo {author} {\bibfnamefont
		{T.}~\bibnamefont {Novotn\'y}},\ }\href {\doibase
	10.1103/PhysRevApplied.11.044094} {\bibfield  {journal} {\bibinfo  {journal}
		{Phys. Rev. Applied}\ }\textbf {\bibinfo {volume} {11}},\ \bibinfo {pages}
	{044094} (\bibinfo {year} {2019})}\BibitemShut {NoStop}%
\bibitem [{\citenamefont {Seth}\ \emph {et~al.}(2016)\citenamefont {Seth},
	\citenamefont {Krivenko}, \citenamefont {Ferrero},\ and\ \citenamefont
	{Parcollet}}]{Seth-2016}%
\BibitemOpen
\bibfield  {author} {\bibinfo {author} {\bibfnamefont {P.}~\bibnamefont
		{Seth}}, \bibinfo {author} {\bibfnamefont {I.}~\bibnamefont {Krivenko}},
	\bibinfo {author} {\bibfnamefont {M.}~\bibnamefont {Ferrero}}, \ and\
	\bibinfo {author} {\bibfnamefont {O.}~\bibnamefont {Parcollet}},\ }\href
{\doibase http://dx.doi.org/10.1016/j.cpc.2015.10.023} {\bibfield  {journal}
	{\bibinfo  {journal} {Comput. Phys. Commun.}\ }\textbf {\bibinfo {volume}
		{200}},\ \bibinfo {pages} {274 } (\bibinfo {year} {2016})}\BibitemShut
{NoStop}%
\bibitem [{\citenamefont {Luitz}\ and\ \citenamefont
	{Assaad}(2010)}]{Luitz-2010}%
\BibitemOpen
\bibfield  {author} {\bibinfo {author} {\bibfnamefont {D.~J.}\ \bibnamefont
		{Luitz}}\ and\ \bibinfo {author} {\bibfnamefont {F.~F.}\ \bibnamefont
		{Assaad}},\ }\href {http://link.aps.org/doi/10.1103/PhysRevB.81.024509}
{\bibfield  {journal} {\bibinfo  {journal} {Phys. Rev. B}\ }\textbf {\bibinfo
		{volume} {81}},\ \bibinfo {pages} {024509} (\bibinfo {year}
	{2010})}\BibitemShut {NoStop}%
\bibitem [{\citenamefont {Pokorn{\'y}}\ and\ \citenamefont {{\v
			Z}onda}(2018)}]{Pokorny-2018}%
\BibitemOpen
\bibfield  {author} {\bibinfo {author} {\bibfnamefont {V.}~\bibnamefont
		{Pokorn{\'y}}}\ and\ \bibinfo {author} {\bibfnamefont {M.}~\bibnamefont {{\v
				Z}onda}},\ }\href {\doibase https://doi.org/10.1016/j.physb.2017.08.059}
{\bibfield  {journal} {\bibinfo  {journal} {Physica B}\ }\textbf {\bibinfo
		{volume} {536}},\ \bibinfo {pages} {488 } (\bibinfo {year}
	{2018})}\BibitemShut {NoStop}%
\end{thebibliography}

%

\end{document}